\documentclass[onefignum,onetabnum]{siamonline250211}
\usepackage{amssymb}
\usepackage{amsmath}
\usepackage{amsfonts}

\usepackage{float}
\usepackage{mwe}
\usepackage{graphbox}
\usepackage{setspace}
\usepackage{epstopdf}
\usepackage{epsf}
\usepackage{subcaption}

\usepackage{bm}
\usepackage{acronym}

\usepackage{lipsum}
\usepackage{graphicx}
\usepackage{algorithmic}
\ifpdf
  \DeclareGraphicsExtensions{.eps,.pdf,.png,.jpg}
\else
  \DeclareGraphicsExtensions{.eps}
\fi

\usepackage{enumitem}
\setlist[enumerate]{leftmargin=.5in}
\setlist[itemize]{leftmargin=.5in}

\newsiamremark{remark}{Remark}
\newsiamremark{hypothesis}{Hypothesis}
\crefname{hypothesis}{Hypothesis}{Hypotheses}
\newsiamthm{claim}{Claim}
\newsiamremark{fact}{Fact}
\crefname{fact}{Fact}{Facts}

\headers{On Inverse Problems, Parameter Estimation, and Domain Generalization}{D. Pereg}

\title{On Inverse Problems, Parameter Estimation, and Domain Generalization}

\author{Deborah Pereg 
				\thanks{The majority of the work was done when the author was with MIT.
				\email{dvorapereg@gmail.com}}}

\usepackage{amsopn}

\begin{document}

\maketitle

\begin{abstract}
Signal restoration and inverse problems are key elements in most real-world data science applications. 
In the past decades, with the emergence of machine learning methods, inversion of measurements has become a popular step in almost all physical applications, normally executed prior to downstream tasks that often involve parameter estimation. 
In this work, we propose a general framework for theoretical analysis of parameter estimation in inverse problem settings.
We distinguish between continuous and discrete parameter estimation, corresponding with regression and classification problems, respectively.
We investigate this setting for invertible and non-invertible degradation processes, with parameter estimation that is executed directly from the observed measurements, comparing with parameter estimation after data-processing performing an inversion of the observations. 
Our theoretical findings align with the well-known information-theoretic data processing inequality, and to a certain degree question the common misconception that data-processing for inversion, based on modern generative models that may often produce outstanding perceptual quality, will necessarily improve the following parameter estimation objective. 
Importantly, by re-formulating the domain-shift problem in direct relation with discrete parameter estimation, we expose a significant vulnerability in current popular practical attempts to enforce domain generalization, which we dubbed the Double Meaning Theorem. 
These theoretical findings are experimentally illustrated 
for domain shift examples in image deblurring and speckle suppression in medical imaging. 
It is our hope that this paper will provide practitioners with deeper insights that may be leveraged in the future for the development of more efficient and informed strategic system planning, critical in safety-sensitive applications.

\end{abstract}

\begin{keywords}
Cram\'er-Rao Bound, Data Processing Inequality, Fisher Information, Inverse Problems, Domain Adaptation.
\end{keywords}

\begin{MSCcodes}
68Q25, 68R10, 68U05
\end{MSCcodes}

\section{Introduction}

Signal restoration and inverse problems have a significant impact on science and engineering applications across numerous disciplines, such as communication, security, healthcare, education and art. 
As restoration algorithms have a fundamental role in signal processing and imaging sciences, there is a growing need to accurately assess their reliability.  
Recent advances in generative artificial intelligence had remarkable impact on the ability of inverse-problem solvers to restore corrupted information and to produce perceptually enhanced results. 
That said, it has been observed that these models are often inclined to generate hallucinations \cite{Alemohammad:2023,Aithal:2024}, which in turn has led to further development of uncertainty quantification methods \cite{Vovk:2005,Romano:2019,Angelopoulos:2022} in an attempt to provide statistical guarantees that could hopefully protect from models' mistakes or hallucinations. Distinct hallucinations were also observed in large language models, e.g., memory distortion, and cognitive mirage, where a model could generate statements that seem reasonable but are
either cognitively irrelevant or factually incorrect \cite{Rawte:2024}.

The recent popularity of deep learning methods have inspired new interest in optimal performance bounds in classification \cite{Jeong:2024} and regression tasks \cite{Shalev:2014}. Machine learning methods have demonstrated outstanding performance for classification tasks. Nevertheless, it is challenging to determine whether such methods have achieved the Bayes error rate, i.e., the lowest error rate obtained by any classifier \cite{Tumer:1996,Martin:1994}. Furthermore, restoration methods (such as denoising, despeckling and deblurring) are increasingly adopted as preprocessing stages prior to downstream tasks, such as classification and segmentation, under the assumption that a restoration pre-processing stage that produces a visually enhanced image, will necessarily improve the results of the following tasks. Whereas, in accordance with the information theoretic data processing inequality \cite{ThomasCover:2006}, we know that every processing stage is bound to lose information.  It is only possible to extract information and it is impossible to recreate new information. 

Most real-world imaging systems acquire measurements that are degraded by
noise, scattering, aberrations, attenuation, and smearing. Conventional cameras' raw sensor measurements are commonly processed in a sequential pipeline, addressing first low-level vision tasks, such as compression, denoising and deblurring, and then proceeding to high-level tasks, such as: classification and feature extraction \cite{Diamond:2021}. In medical imaging, it is commonly assumed that image-enhancement pre-processing stages (e.g., denoising, deblurring and super-resolution) would necessarily boost the performance of post-processing parameter-estimation downstream stages. For example, it is often assumed that speckle suppression of ultrasound and optical coherence tomography (OCT) images should be employed prior to parameter estimation, such as segmentation and elasticity assessment to determine the progress of glaucoma \cite{Pereg:2023B,Pereg:2024A,Braeu:2024,Yan:2020}. 

The term class imbalance refers to clinical prediction models with a binary outcome where the event occurrence is much lower than $50\%$
imbalance correction methods, such as: random undersampling, random oversampling, or synthetic minority oversampling technique (SMOTE) led to models with strong miscalibration without better ability to distinguish
between patients with and without the outcome event. Clearly, inaccurate probability estimates yield ill-informed clinical decisions, thus reducing the utility of the prediction models.
Since the accuracy measure of a classifier depends on the percentage of the true positives and true negatives, a trained classifier may appear to have a high score when the majority of events belong to one class, but the minority class, which is often the high risk class, is misclassified. Similar results were presented for machine learning based prediction models \cite{Carriero:2024}.

In the past, most signal restoration algorithms were deterministic, that is, they were designed to provide one restored
signal given a degraded input measurement (or several input measurements). 
Recent stochastic restoration algorithms \cite{Song:2020, Simoncelli:2021, Ohayon:2023a, Ohayon:2023b, Milanfar:2023}
explore the possible solutions by sampling from a distribution that is conditioned on the degraded input. 
This approach  applies the inherent nature of most of these ill-conditioned inverse problems, that is - there is more than one solution that can fit the measurement, and there is an inherent uncertainty in real-data applications regarding which solution should be favored.
That said, apart from the ability of stochastic approaches to explore different viable solutions, it is still unclear what is their advantage in terms of perceptual quality and robustness. In the context of this work an interesting question is whether stochastic restoration methods can improve the following parameter estimation. 

Recent experimental studies observe that image-to-image restoration models are extremely sensitive to adversarial attacks, which in certain cases is diagnosed as mode-collapse.
Namely, even visually undetectable input perturbations lead to severe output artifacts \cite{Choi:2019,Choi:2022,Yan:2022}.
In this work, we establish an alternative reasoning behind this significant vulnerability. 
We prove that training over several domains in an attempt to improve domain generalization (Double Meaning Theorem (Theorem \ref{theorem8})) may lead to ambiguity in the required output, which in turn inevitably leads to a degraded output. 
Our theoretical findings are experimentally illustrated for image deblurring and speckle supression. 

The general objective of this research work is to provide a mathematical analysis along with the corresponding intuition for the exploration of the optimal path towards parameter estimation in the context of inverse problems. To this end, we analyze several specific settings. 
In Section~\ref{Sec3}, we prove a fundamental property of domain generalization (Double Meaning Theorem (Theorem \ref{theorem8})). We empirically validate our theoretical findings via experiments in natural image deblurring and medical image denoising. In Section~\ref{Sec4}, we introduce a general problem formulation for parameter estimation in an inverse problem setting, and define a perfect perception estimator. We derive fundamental theorems for Bayes error bounds in classification problems following preprocessing of acquired measurements (Theorems~\ref{theorem5}-\ref{theorem7}), and provide illustrative examples to inspire intuition into the proposed point-of-view.  
Lastly, in Section~\ref{Sec5}, we prove that in a continuous setting a perfect perception estimator cannot improve the accuracy of the following parameter estimation, whether it is conditioned on that parameter either explicitly or implicitly, but it may achieve equivalent performance (Theorems \ref{theorem2}-\ref{theorem4}). 

We note that a substantial portion of this paper may be viewed as a review of existing knowledge, and is included for the sake of completeness. We strongly believe that this work emphasizes important aspects for practitioners in real-world applications. 

\newpage
\section{Problem Setting}\label{Sec2}

\subsection{Notations}
Let $X \in \mathcal{X}^{n\times 1}$ and $Y \in \mathcal{Y}^{d\times 1}$ be random variables
that obey stationary and ergodic probability distributions, and have a stationary coupling \cite{Gray:2009}.
Throughout the majority of this work, we consider a setting where $X$ and $Y$ are continuous
random variables. 
It is also possible to consider a practical setting where $X$ and $Y$ are continuous
random variables represented in a ﬁnite precision machine where both $X$ and $Y$ are quantized into a ﬁnite number of
discrete values. In the discrete case the proofs generally follow similar outline where the corresponding integrals should be replaced with finite sums.  
Note that here $Y$ may be a single measurement or a vector of $m\in \mathbb{N}$ measurements. 
We denote the joint probability of $X$ and $Y$ as $p_{X,Y}(x,y)$, and their corresponding mutual information
is deﬁned as $I(X;Y) =\mathcal{D}(p(x,y)||p(x)p(y))\geq0$, where we have used the standard notation
$\mathcal{D}(p||q) \triangleq \int p(u) \log \frac{p(u)}{q(u)} du$ for the Kullback-Liebler (KL) divergence between the probability density functions $p$
and $q$. Namely, in a continuous setting,
\begin{equation}\label{1.0}
I(X;Y)=\int p(x,y) \mathrm{log} \frac{p(x,y)}{p(x)p(y)} dx \ dy.
\end{equation}

Equivalently, we assume a sample space that is a set $\Psi$ of paired objects $\Psi=\{\mathbf{y}_{i},\mathbf{x}_{i}\}_{i=1}^m$, where $\mathbf{x}_i \in  \mathcal{X}^{n\times 1}$ are sampled from $p_{\theta}(\mathbf{x})$ and paired with $\mathbf{y}_i \in  \mathcal{Y}^{d\times 1}$ by a deterministic or stochastic mapping as ground truth. $\Psi_x=\{\mathbf{x}_i\}_{i=1}^{m}$, can be thought of as possible outcomes of a ``perfect" experiment distributed according to an unknown distribution of a certain set of probability measures that depend on the true parameter $\theta$. Whereas $\Psi_y=\{\mathbf{y}_i\}_{i=1}^{m}$ are the corresponding measured (degraded) outcomes. For example, one the most familiar concrete examples in statistics \cite{Fisher:1922,Halmos:1949} is an $n$-dimensional vector $\mathbf{x} \in\mathbb{R}^{n \times 1}$, of $n$ element-values representing $n$ independent observations of a normally
distributed random variable with unknown parameters.
The notations $\mathbf{x}$ and $X$ are used below interchangeably. 

When $p(x|\theta)$ is viewed as a function of $x$ with 
$\theta$  fixed, it is a probability density function (pdf), and when it is viewed as a function of 
$\theta$  with $x$ fixed, it is a likelihood function. 
The notation $p_\theta(x) \triangleq p(x|\theta)$ indicates that $\theta$ is regarded as a fixed unknown quantity, rather than as a random variable being conditioned on.

\subsection{Ergodic Sources}

Let $x^n$ denote a sequence $x_1,x_2,...,x_n$, and $\mathbf{x}\in\mathcal{X}^{n \times 1}$ to denote a vector with $n$ entries. In information theory, a stationary stochastic process $u^n$ taking values in some finite alphabet $\mathcal{U}$ is referred to as a source. More often than not, communication theory refers to discrete memoryless sources (DMS) \cite{Kramer:2008,ThomasCover:2006}. That said, many signals, such as image patches, are usually modeled as entities belonging to some probability distribution forming statistical dependencies (e.g., a Markov random field (MRF) \cite{Roth:2009,Weiss:2007}) corresponding to the relations between data points in close spatial or temporal proximity.
Although the formal definition of ergodic processes is somewhat complicated (\cite{Gray:2009,Papoulis:1991}, Appendix \ref{appA}), as described by Shannon in 1948, the general idea is simple: ``In an ergodic process, every sequence that is produced by the process is the same in statistical properties" \cite{Shannon:1948}. Particular sequences generated by the process will exhibit symbol frequencies approaching a definite statistical limit, as the lengths of the sequences is increased.
More formally, we assume an ergodic source with memory that emits $n$ symbols from a discrete and finite alphabet $\mathcal{U}$, with probability $P_U(u_1,u_2,...,u_n)$.
We recall a theorem \cite{Breiman:1957}, here without proof.
\begin{theorem}
[Entropy and Ergodic Theory \cite{Breiman:1957}]%
\label{theorem1} Let $(u_n)_{n \in \mathbb{Z}}$ be a stationary ergodic process ranging over a finite alphabet $\mathcal{U}$, then there is a constant $H$, defining the entropy rate of the source,
\begin{equation*} 
H = \lim_{n\rightarrow\infty} -\frac{1}{n} \log_2 P_U(u_1,...,u_n).
\end{equation*} 
\end{theorem}

Intuitively, when we observe a source with memory over several time units, once we know the previous source's entries, the dependencies reduce the overall conditional uncertainty, thus the uncertainty grows more slowly as $n$ grows. 
The entropy rate $H$ representing the average uncertainty per time unit, converges over time. 
Although the entropy rate is typically used for discrete memoryless sources (DMS), 
every ergodic source has the Asymptotic Equipartition Property (AEP) \cite{Breiman:1953}. 
The generalization of the AEP to arbitrary ergodic sources can be found in \cite{Breiman:1953}.
Entropy typicality applies also to continuous random variables with a density $p_U$ replacing the discrete probability $P_U(u^n)$ with the density value $p_U(u^n)$. 
The AEP leads to Shannon's source coding theorem stating that the average number of bits required to specify a symbol in a sequence of length $n$, when we consider only the most probable sequences, is $H$.
And it is the foundation for the known rate-distortion theory and channel capacity. 
In Pereg et al. (2023) \cite{Pereg:2023A}, we took a similar stand-point with respect to ergodic sources to apply the AEP to bound the sample complexity for supervised learning in regression models with respect to the empirical error and the generalization error.

\subsection{Fisher information}

We assume $\theta$ is an unknown deterministic true parameter that we wish to estimate, 
given a set of $m$ independent observations (measurements) of a signal 
$\{\mathbf{x}_i\}_{i=1}^{m}:\mathbf{x}_i \sim p_{\theta}(\mathbf{x})$.
We define the score as
\begin{equation}\label{1.1}
s(\theta) = \frac{\partial}{\partial\theta}\log p_{\theta}(\mathbf{x}).
\end{equation}
Note that $E\{s(\theta)|\theta\}=0$, where $E$ denotes mathematical expectation.

Fisher information (FI) is defined as the variance of the score \cite{Fisher:1922},
\begin{equation}\label{1.2}
J_{X}(\theta)=E\Bigg\{\Bigg( \frac{\partial}{\partial\theta}\log p_{\theta}(\mathbf{x}) \Bigg)^2 \Bigg| \theta \Bigg\}.
\end{equation}
$J_{X}(\theta) \geq 0$. The FI is defined only under regulatory conditions of differentiability and integrability.
For example, a simple case where the Cram\'er-Rao bound may not be valid is when the support of $p_{\theta}(\mathbf{x})$ depends on $\theta$. 
If we consider a sample of $m$ random variables $X_1,X_2,...,X_m$ drawn i.i.d. from $p_{\theta}(x)$, we have
$p_{\theta}(x_1,x_1,...,x_m)=\Pi_{i=1}^{m} p_{\theta}(x_i)$ and 
\begin{equation*}
J^m_{X}(\theta)= m J_{X}(\theta)
\end{equation*}
Notably, the chain rule for the FI \cite{Zamir:1998} is
\begin{equation*}
J_{X,Y}(\theta)=J_{X}(\theta)+J_{Y|X}(\theta)
\end{equation*}

\subsection{Cram\'er-Rao Bound}
Considering an unbiased estimator $\hat{\theta}_X$ extracted from the observed realization of a random variable $X$, such that
\begin{equation}\label{1.3}
B=\theta-E\hat{\theta}_X=\int \big(\hat{\theta}_X-\theta\big)p_\theta(x)dx=0  \quad \forall \theta,
\end{equation}
The variance of any unbiased estimator $\hat{\theta}_X$ of $\theta$ is bounded by the reciprocal of Fisher information $J_X(\theta)$,
\begin{equation}\label{1.4}
E(\theta-\hat{\theta}_X)^2 \geq \frac{1}{J_X(\theta)},
\end{equation}
In other words, the Fisher information of the likelihood function fundamentally bounds the precision to which we can estimate $\theta$, for unbiased estimators.

Additional relevant background and definitions can be found in Appendix \ref{appA}.

\subsection{Problem Formulation}
Let $Y$ be an observation, and $\hat{\theta}_Y$ be the estimated parameter given $m$ examples (realizations) of $Y$. We assume $\hat{X}=\mathcal{F}(Y)$ is an estimated signal that is derived from $Y$ by solving an inverse problem, in an attempt to recover an underlying source signal $X$, and $\hat{\theta}_{\hat{X}}$ as the estimated parameter given $m$ examples of $\hat{X}$. In other words, $\theta-X-Y$ is a Markov chain.
That is, given $X$, $Y$ and $\theta$ are independent. Namely, $p_{\theta}(x,y) \triangleq p(x,y|\theta)=p_\theta(x)p(y|x)$
(i.e., the conditional
distribution of $Y$ given $X$ is independent of $\theta$). \\
In addition,
$\theta-X-Y-\hat{\theta}_Y$ is a Markov chain, and
$\theta-X-Y-\hat{X}-\hat{\theta}_{\hat{X}}$ is also a Markov chain.
\begin{definition}
A degradation $p_{Y|X} (\cdot|x)$ is invertible if the conditional probability distribution $P_{X|Y} (\cdot|y)$ is a
delta measure for almost every $y$.
\end{definition}
In other words, given the observation $Y=y$, the original signal $X$ is uniquely determined (almost surely).
A degradation is invertible if there exists a measurable function $\mathcal{G}: \mathcal{Y}\rightarrow \mathcal{X}$ such that 
$Pr[X=\mathcal{G}(Y)]=1$.

\textbf{Remark 1.} Note that the above problem setting is somewhat ambiguous. One may ask whether we have access to $m$ realizations of $Y$ given a specific $\theta$, or do we have $m$ examples of $Y$ for different $\theta$'s. 
We will see that this distinction is indeed significant. 

\begin{figure}
\includegraphics[width=10.5cm, height=7.3cm]{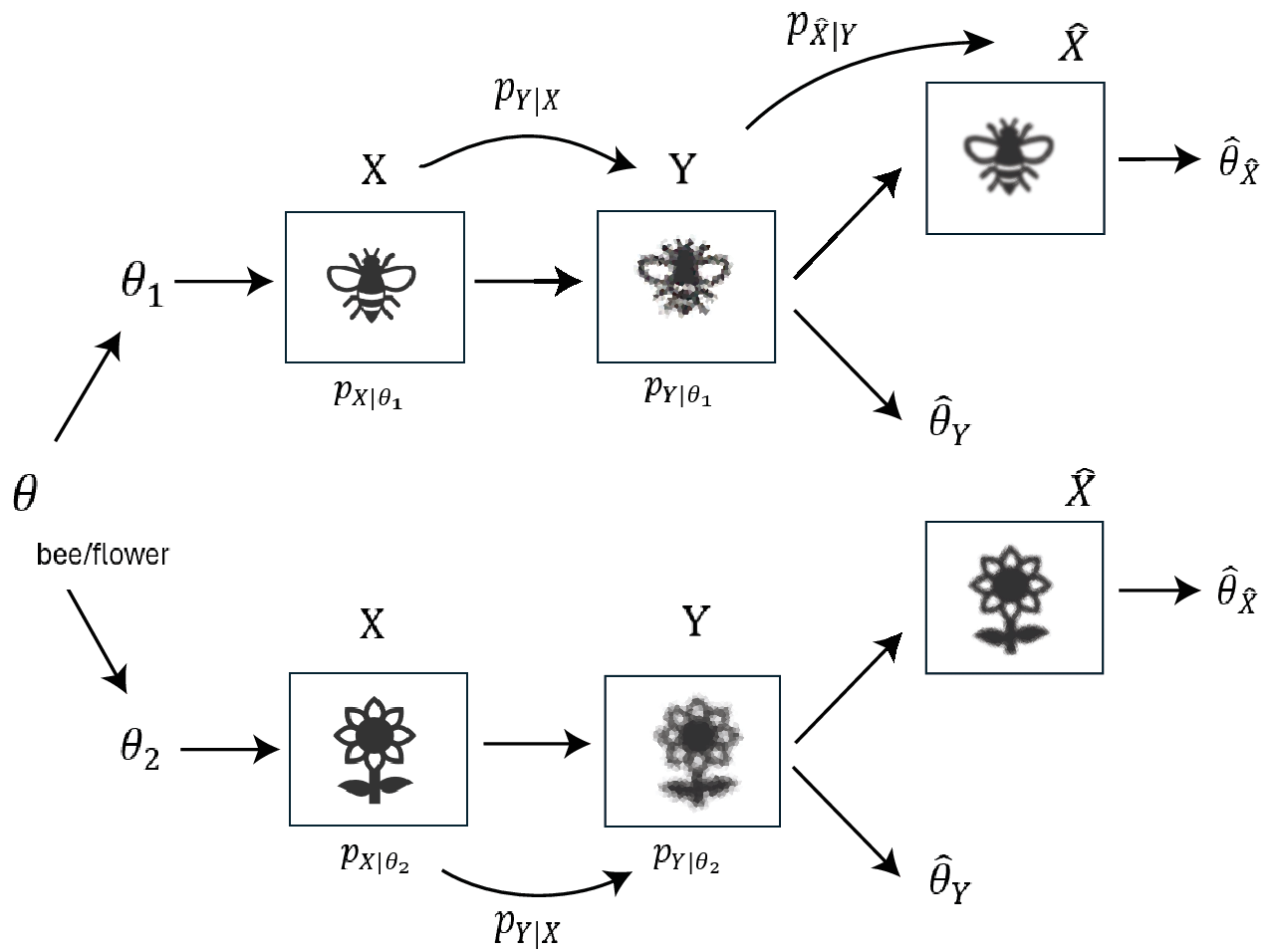}
\caption{Illustration of an example of the problem setting: Classification with or without image-to-image regression.
We observe a degraded image $y \sim p_{Y|\theta}$ according to some conditional probability density function $p_{Y|X}$, where the original image is a realization from a probability density function $x\sim p_{X|\theta}$.}
\label{fig1a}
\end{figure}

\begin{figure}
\includegraphics[width=10.5cm, height=4cm]{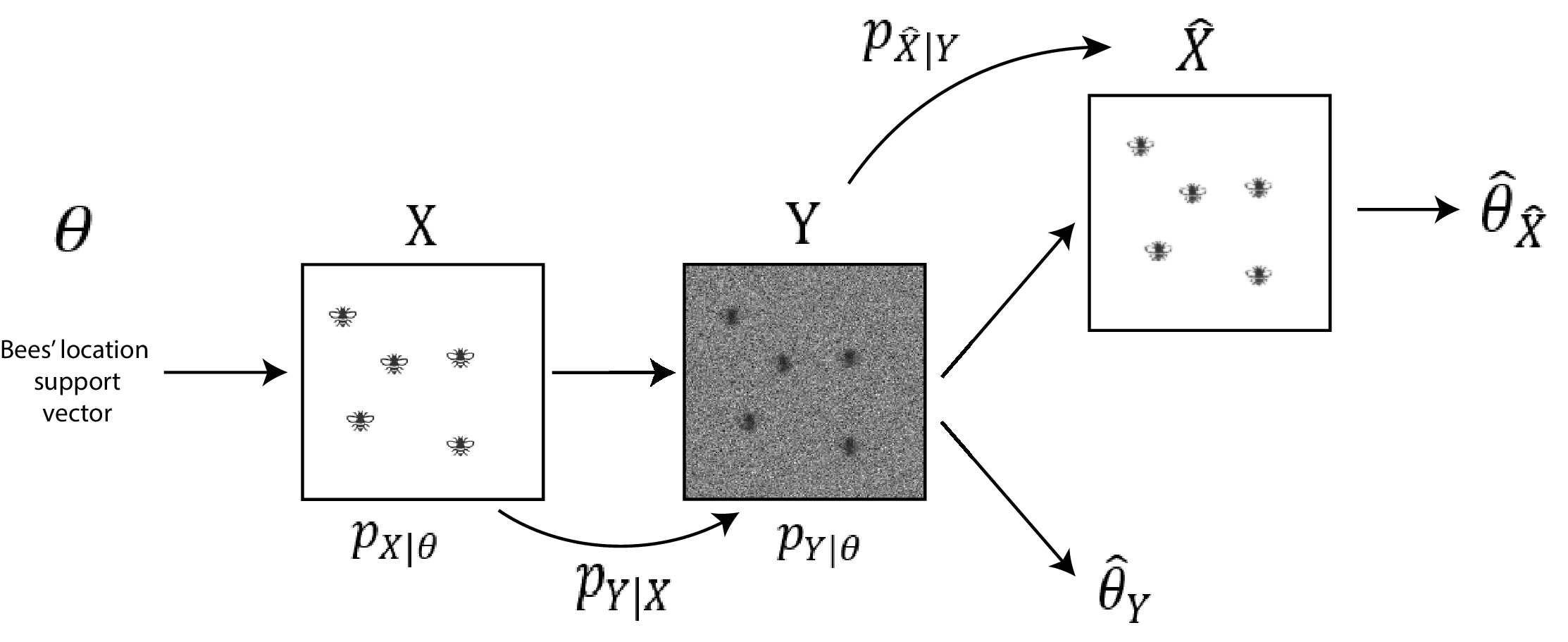}
\caption{Illustration of an example of the problem setting: Sparse support detection with or without denoising and deblurring.}
\label{fig1b}
\end{figure}

Figure \ref{fig1a} illustrates a simplified example of the problem setting for a regression image-to-image denoising (inversion) followed by a classification where $\theta$ is a discrete variable representing one of two classes. Figure \ref{fig1b} illustrates a similar image restoration regression problem followed by sparse support detection where $\theta$ is a sparse support location vector of a known object. Clearly, reconstructing the source image would provide the observer with a nice-looking image where it is easier for a human observer to solve the following task. But given that the pre-processing stage is erroneous, it is obvious that an imperfect reconstruction would inevitably lead to the propagation of its errors into the following stages. 

\newpage
\section{Towards mathematical theory of domain variation}\label{Sec3}
Throughout this section, we consider a discrete setting where $X \in \mathcal{X}^{n\times 1}$ and $Y \in \mathcal{Y}^{d\times 1}$ are discrete random variables.
\begin{definition}
In a discrete setting, a degradation $p_{Y|X} (\cdot|x)$ is invertible if the conditional probability mass function $p_{X|Y} (\cdot|y)$ is a
delta Kronecker for all $y \in \mathrm{supp}(p_Y)$.
\end{definition}
\textbf{When M classes represent different domains.} \\
Consider a mixture model $p(\mathbf{x})=\sum^M_{i=1}P(\theta_i)p(\mathbf{x}|\theta_i)$, 
and $p(\mathbf{y}|\mathbf{x})$ independent of $\theta$. 
Therefore, $p(\mathbf{y})=\sum^M_{i=1}P(\theta_i)p(\mathbf{y}|\theta_i)$.
Note that, even when $\mathbf{y}|\theta_i=T_{\theta_i}(\mathbf{x}) \forall i$ is an invertible mapping, unless the classes $p(\mathbf{y}|\theta_i)$ have disjoint supports, $p(\mathbf{y})$ is no longer invertible. 

\textbf{The Domain-Shift Problem.} 
In learning theory, domain shift is a variation of the data distribution between the source domain (training dataset) and
target domain (the test dataset). Supervised learning models have demonstrated remarkable results in image-to-image translation tasks across numerous applications \cite{Elad:2023,Pereg:2023B,Pereg:2024B}. 
However, it has been observed that while sufficient results were obtained for the training domain (source domain), these models collapse when tested on unseen target domains, such as under contrast and resolution variations \cite{Pan:2021}. One of the most common techniques to improve generalization is data augmentation, normally accomplished by applying simple spatial and intensity transforms to the training data.
Another popular strategy to improve generalization is domain randomization \cite{Tobin:2017} attempting to significantly enlarge the training distribution beyond the source domain, under the hypothesis that learning from out-of-distribution examples improves generalization \cite{Bengio:2011}.
Most domain adaptation and transfer learning methods operate under the assumption that there exists a shared hypothesis that performs well in both domains \cite{Ben:2010}. 
Here, we argue that certain implementations of this strategy may lead to degraded results, because instead of learning out-of-distribution examples (extrapolation), the model learns new examples within the distribution's support that compete with the desired task and create ambiguity, which in turn leads to mode-collapse. 

\subsection{Signal Restoration over Several Domains}
For the sake of simplicity, let us assume $M$ domains such that each noisy and/or degraded measurement $\mathbf{y}\in\mathcal{Y}^n$ originates in a \textit{different} true signal in each domain $\{\mathbf{x}_i\}_{i=1}^M$. Namely,  
\begin{equation}
p_{\theta_i}(\mathbf{x}_i|\mathbf{y}) = \delta(\mathbf{x}_i-G_i(\mathbf{y})), \  G_i(\mathbf{y}) \neq G_j(\mathbf{y}) \ \ \forall i \neq j, \ i,j=1,...,M,  
\end{equation}
$G_i(\mathbf{y})$ is a deterministic function. 
In other words, in different domains the forward model $\mathcal{T} : \mathbf{x} \rightarrow \mathbf{y}$ (random or deterministic) leads different source signals to the same observed measurement.
In different domains the observed measurement originates in different source signals.
We are training an algorithm in a supervised fashion to produce a deterministic mapping $\mathcal{F} : \mathbf{y} \rightarrow \mathbf{x}$ such that $p_{\hat{X}|Y}(\hat{\mathbf{x}}|\mathbf{y})=\delta\big(\mathcal{F}(\mathbf{y})-\hat{\mathbf{x}}\big)$.
When adopting domain randomization approach 
\footnote{We note that domain randomization inherently assumes that all domains are equally probable. 
That is, $P(\theta_i)=\frac{1}{M} \forall \ i$.
And one may wonder if this is indeed the best approach?}, or a data augmentation approach \footnote{Under a data augmentation strategy that generates examples from domains out of the source domain.},
we are optimizing a neural net's parameters by minimizing some loss function
$\mathcal{L}(\hat{\mathbf{x}},\mathbf{x}_i)$ 
that has the smallest average deviation from the expected outputs.
But since we have $M$ domains, i.e., $M$ valid outputs, we now have
\begin{equation}
\arg\min_{\hat{\mathbf{x}}}
\sum_{i=1}^{M}{\ell(\hat{\mathbf{x}},\mathbf{x}_i)},
\end{equation}
where 
$\ell(\hat{\mathbf{x}},\mathbf{x})$ is the loss (empirical risk). 
For the MSE loss $\mathcal{L}(\mathbf{x},\mathbf{x}_i) = \sum_{i=1}^{M}(\mathbf{x}-\mathbf{x}_i)^2$
the minimum is obtained at the arithmetic mean of the possible outputs,
\begin{equation}
\hat{\mathbf{x}}=
\frac{1}{M} \sum_{i=1}^{M}\mathbf{x}_i.
\end{equation}
For the $\ell_1$ loss, the sum of absolute deviations $\ell(\hat{\mathbf{x}},\mathbf{x}_i) = |\hat{\mathbf{x}} -
\mathbf{x}_i|$, the optimum is at the median of the expected outputs. This phenomenon is also referred to in the literature as ``regression to the mean" \cite{Elad:2023}, but normally in the context of having multiple restoration solutions in one domain due to the ill-posed nature of the inversion task (see also Section~\ref{Sec4.3.2} below). 
Here, we emphasize that even if the restoration function is invertible in each domain 
\footnote{That is, in each domain the posterior is a delta function with nonzero probability for only one
reconstruction. Therefore it is possible to perfectly reconstruct the signal from the measurement.} 
unless the sets are disjoint, 
domain randomization would directly lead to averaging over the possible outputs.
Let us summarize the above conclusion in the following theorem,
\begin{theorem}[Double Meaning Theorem]\label{theorem8}
Let $p(\mathbf{x}|\theta_i), \ i=1,...,M$ be M domains, and $p(\mathbf{y}|{\mathbf{x}})$ be a degradation process, such that the observation sets are not disjoint, i.e., $p_{\theta_i}(\mathbf{x}|\mathbf{y})$ have joint supports in the $\mathbf{y}$ input domain.
Let $\mathcal{F}(\mathbf{y})$ be an inversion model that is successfully trained by domain randomization. That is, the model is trained to 
minimize a cost function over all given and equiprobable domains, $P(\theta_i)=\frac{1}{M}, \ \forall i=1,...,M$,
such that the model $\mathcal{F}$ is trained by minimizing the empirical loss 
\begin{equation}
\mathcal{R}_m(\mathcal{F}) = \frac{1}{m} \sum_{k=1}^m \ell(\mathcal{F}(\mathbf{y}_k),\mathbf{x}_k). 
\end{equation}
where $m$ is the sample size. 
For $m$ large enough, the inversion output yields the minimal average expected loss over the possible outputs in the different domains,
\begin{equation}
\hat{\mathbf{x}}= \mathcal{F}(\mathbf{y}) =
\arg\min_{\hat{\mathbf{x}}}
\frac{1}{M} \sum_{i=1}^{M}E_{\mathbf{x}_i\sim\mathbf{x}|\mathbf{y},\theta_i}\big[\ell(\hat{\mathbf{x}},\mathbf{x}_i)\big],
\end{equation}
If $p(\mathbf{x}|\mathbf{y},\theta_i)$ collapses to a Dirac delta at $\mathbf{x}_i$ for each domain $i$, then we have
\begin{equation}
\hat{\mathbf{x}}= \mathcal{F}(\mathbf{y}) =
\arg\min_{\hat{\mathbf{x}}}
\frac{1}{M} \sum_{i=1}^{M}{\ell(\hat{\mathbf{x}},\mathbf{x}_i)}.
\end{equation}
\end{theorem}
\textit{Proof}. See Appendix~\ref{appC}. \\
For example, suppose we have $M=2$ possible domains such that $p_{\theta_1}(\mathbf{x}_{\theta_1}|\mathbf{y}) = \delta(\mathbf{x}_{\theta_1}-G_1(\mathbf{y}))$, and $p_{\theta_2}(\mathbf{x}_{\theta_2}|\mathbf{y}) = \delta(\mathbf{x}_{\theta_2}-G_2(\mathbf{y}))$, and an MSE error, then we have $\hat{\mathbf{x}}(\mathbf{y})= \frac{1}{2}\mathbf{x}_{\theta_1}+\frac{1}{2}\mathbf{x}_{\theta_2}$

\textbf{Remark 2.} Note that when the sets $\{\mathbf{y} \sim p_{\theta_i}(\mathbf{y})\}_{i=1}^{M}$ are disjoint, it is possible to train a model in a randomized manner over these distributions to generalize over the general $p(\mathbf{x}|\mathbf{y})$ without yielding averaged results. This may be the case for certain implementations of segmentation over different domains \cite{Dey:2024}, and for the MNIST dataset that can be modeled as a Gaussiam mixture model, under the assumption that the Gaussians are ``far enough" from each other (in which case the supports are approximately disjoint).

\textbf{Remark 3.} We emphasize here that the domain randomization and data augmentation strategies are inherently averaging strategies that optimize different scores across the given domains, but not necessarily the score for each domain separately. The outcome is therefore an averaged output across the domains, that may yield impressive quantitative scores on average, yet in practice could often result in reconstructed signals that are smeared, degraded, and/or inconsistent with the input's domain, which could be expressed in incorrect features, such as: contrast, sampling rate, artifacts, noise, and more. 

\textbf{Remark 4.} A similar proof outline applies also when a model is trained with noisy or perturbed training data (which can be treated as an adversarial attack \cite{Sofer:2025}) and for other regression tasks. In those cases the model converges to the mean, which, depending on the task is either beneficial or misleading (see Section~\ref{Sec3D} below, and \cite{Pereg:2026}). 

\textbf{Simple is beautiful.}
In previous works, it has also been observed that synthetic data training may improve generalization \cite{Pereg:2020,Dey:2024,Pereg:2024B}. In those works, the synthetic data is of simpler visual form, almost degenerated. This may be explained by the opposing approach. That is, one may claim that using synthetic data of  extremely simplified forms we are significantly limiting the learned distribution, and rather than expanding it, we are contracting the example set which the model has seen and is able to recognize. And so, by accurately learning simpler representations - the general result is improved. The model learns the basic pattern that may align with many distributions, inherently learning the overlapping support shared across a greater number of possible varying distributions.
An alternative theoretical explanation relying on the information theoretic method of types is presented in \cite{Pereg:2023A}.

We now provide an illustration of the core idea in the Double Meaning Theorem (Theorem \ref{theorem8}) in two relatively simple yet pragmatic examples. 

\begin{figure}[t!]
    \begin{subfigure}[t]{0.16\textwidth}
        \includegraphics[width=0.95\linewidth]{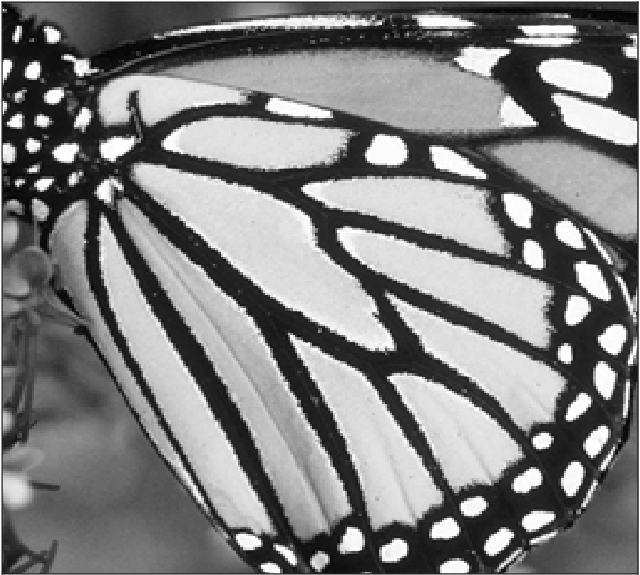}
				\caption{}
    \end{subfigure}%
		\hfill 
		\begin{subfigure}[t]{0.16\textwidth}
        \includegraphics[width=0.95\linewidth]{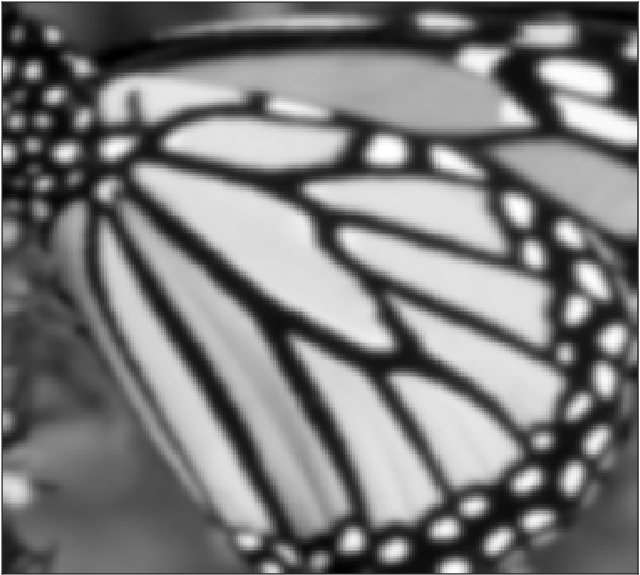}
				\caption{}
    \end{subfigure}%
		\hfill 
    \begin{subfigure}[t]{0.16\textwidth}
        \includegraphics[width=0.95\linewidth]{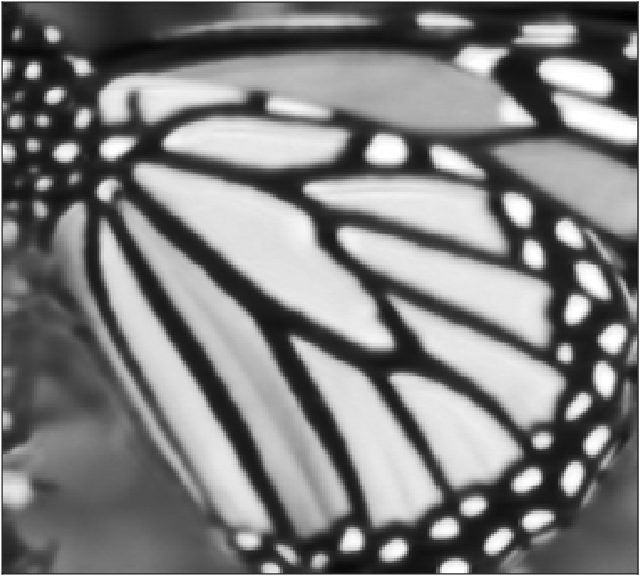}
				\caption{}
    \end{subfigure}%
		\hfill 
		\begin{subfigure}[t]{0.16\textwidth}
        \includegraphics[width=0.95\linewidth]{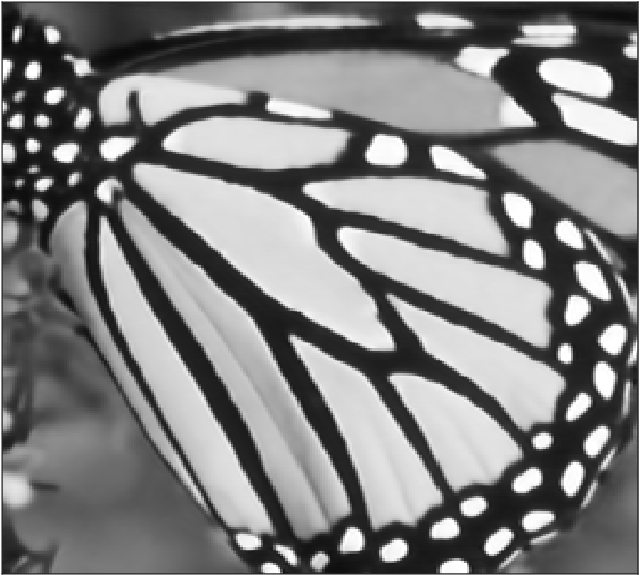}
				\caption{}
    \end{subfigure}%
		\hfill
		\begin{subfigure}[t]{0.16\textwidth}
				\includegraphics[width=0.95\linewidth]{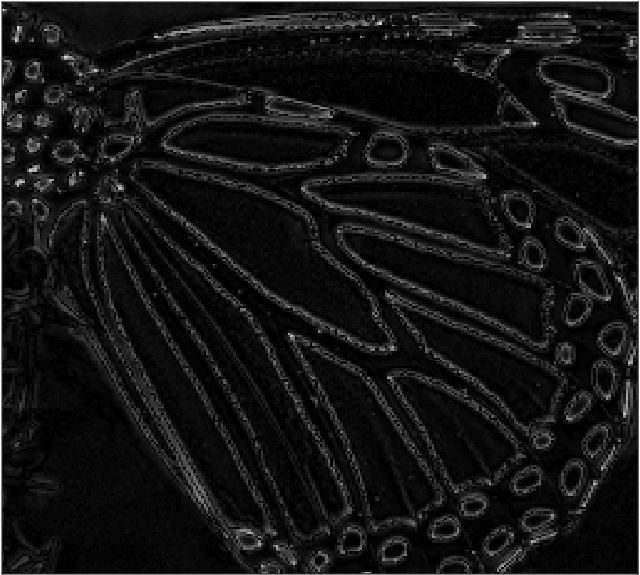}
				\caption{}
		\end{subfigure}%
		\hfill
		\begin{subfigure}[t]{0.16\textwidth}
				\includegraphics[width=0.95\linewidth]{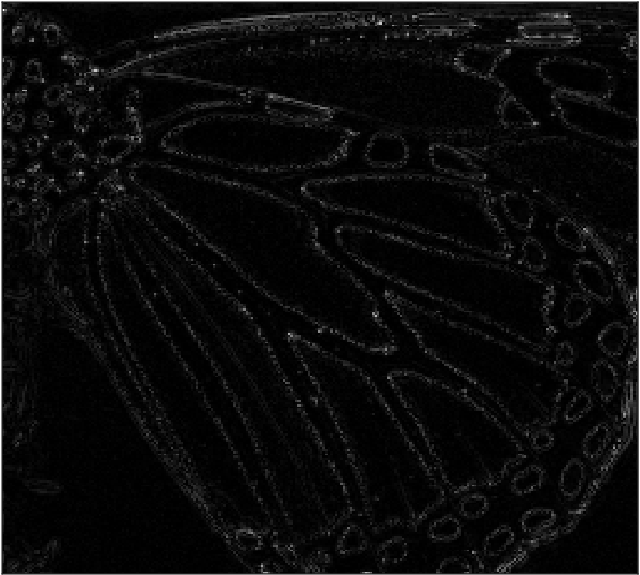}
				\caption{}
    \end{subfigure}%
\caption{{\small Visual comparison of deblurring of the image butterfly: (a) Ground truth; (b) Degraded input; (c) RNN-GAN with mixed training;
(d) RNN-GAN with targeted training; (e) Residual for mixed training; (f) Residual for targeted training.}}
\label{fig6}%
\end{figure}

\subsection{Example 1 - Resolution Shift}

Assume a source image and a degraded image in vector form, $\mathbf{x}, \mathbf{y} \in \mathbb{R}^{N \times 1}$, 
such that,
$\{\mathbf{x},\mathbf{y} \sim p_{\theta_i}(\mathbf{x},\mathbf{y}) : \mathbf{y}=\mathbf{H}_i\mathbf{x}+\mathbf{w}, \ i=\{1,2\}\}$.
$\mathbf{H}_i$ is the convolution matrix of a Gaussian filter with standard deviation $\sigma_i$, where $\sigma_2 = 2\sigma_1=2$, and $\mathbf{w}$ is additional white Gaussian noise (WGN) of level $\sigma_n=\sqrt{2}$. 
To validate our theoretical results, we compare trained one-shot supervised learning models for the task of image deblurring \cite{Pereg:2023B,Pereg:2024B}. 
Recall that for a Gaussian PSF, the convolution of two Gaussian with mean $\mu_1, \mu_2$ and variance $\sigma^2_1$, $\sigma^2_2$ is a Gaussian with mean $\mu=\mu_1+\mu_2$ and variance $\sigma^2 =\sigma^2_1+\sigma^2_2$.
Therefore, under this model, there exist $\mathbf{x}_1 \neq \mathbf{x}_2$ such that $\mathbf{y}=\mathbf{H}_1\mathbf{x}_1=\mathbf{H}_2\mathbf{x}_2$. Let us denote $\mathbf{y}_1=\mathbf{H}_1\mathbf{x}_1, \ \mathbf{y}_2=\mathbf{H}_2\mathbf{x}_2$, where we have omitted the noise term for that sake of simplicity. 
An identical degraded input image obeys $\mathbf{y}_1=\mathbf{y}_2$. Therefore,
\begin{equation}
\mathbf{x}_1*\mathbf{h}_1 = \mathbf{x}_2*\mathbf{h}_2= \mathbf{x}_2*\mathbf{h}_1*\mathbf{h}_{1\rightarrow 2},
\end{equation}
where $\mathbf{h}_{1\rightarrow 2}$ is a Gaussian filter with $\sigma^2_{1 \rightarrow 2}=\sigma_2^2-\sigma_1^2$. 
That is, $\mathbf{x}_1 = \mathbf{x}_2*\mathbf{h}_{1\rightarrow 2}$. 
Consequently, according to Theorem~\ref{theorem8}, the outcome of a model that optimizes the MSE over both outcomes and
thus averaging over both viable reconstructions is,
\begin{equation}
\hat{\mathbf{x}}= \frac{1}{2} \mathbf{x}_1 + \frac{1}{2} \mathbf{x}_2 = \frac{1}{2} \mathbf{x}_2 + \frac{1}{2} \mathbf{x}_2 * \mathbf{h}_{1\rightarrow 2} = \frac{1}{2} [\mathbf{I+H}_{1\rightarrow 2}] \mathbf{x}_2 .
\end{equation}
Note that for some practical examples the result can visually seem improved due to the visual dominance of the first term.
Also, in some applications, or where $M \rightarrow \infty$ an averaged solution would misleadingly appear visually acceptable as a sufficient reconstruction. 

\begin{figure}[t!]
\hspace{2.5cm}
    \begin{subfigure}[t]{0.16\textwidth}
        \includegraphics[width=0.95\linewidth]{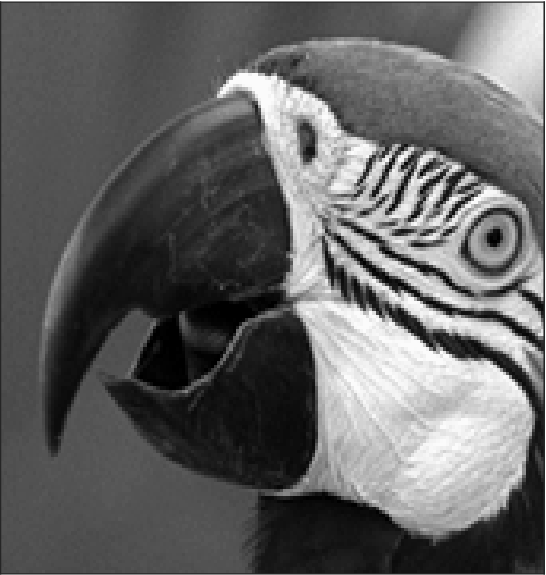}
				\caption{}
    \end{subfigure}
		\begin{subfigure}[t]{0.16\textwidth}
        \includegraphics[width=0.95\linewidth]{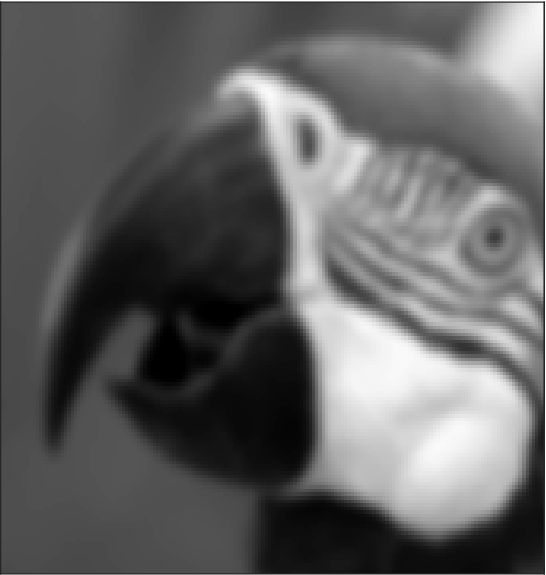}
				\caption{}
    \end{subfigure}
    \begin{subfigure}[t]{0.16\textwidth}
        \includegraphics[width=0.95\linewidth]{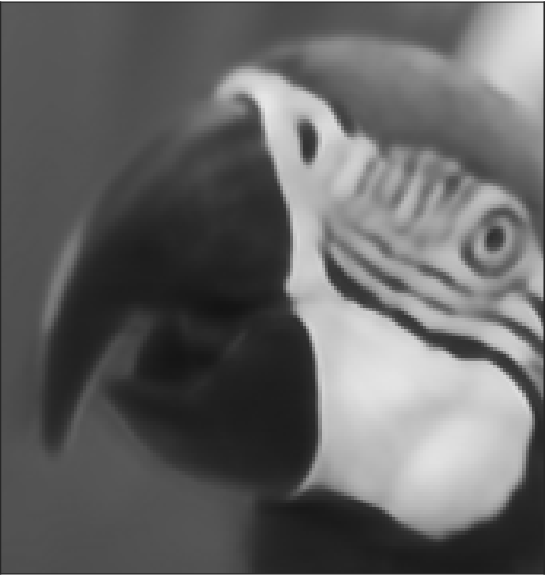}
				\caption{}
    \end{subfigure} 
		\begin{subfigure}[t]{0.16\textwidth}
        \includegraphics[width=0.95\linewidth]{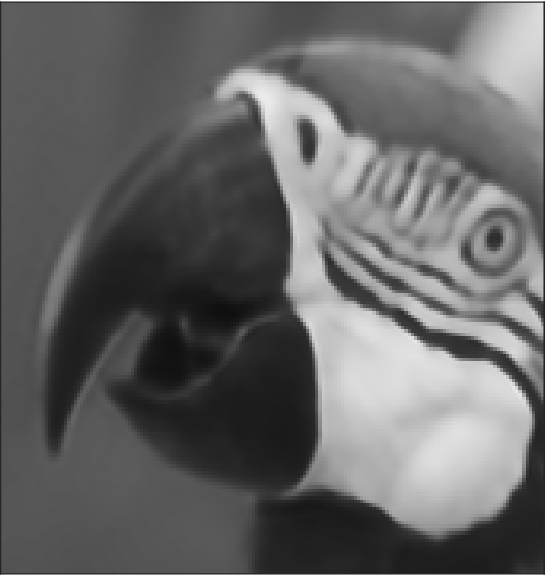}
				\caption{}
    \end{subfigure}
\caption{{\small Visual comparison of deblurring of the image parrot: (a) Ground truth; (b) Degraded Input; (c) RNN-GAN with mixed training;
(d) RNN-GAN with targeted training. Note the changes in the patterns on the parrot's beak and the shadow artifact under the beak with mixed training.}}%
\label{fig7}%
\end{figure}

To illustrate the Double Meaning theorem above, we present a simple example using a one-shot training framework previously described in \cite{Pereg:2023B,Pereg:2024B}. 
Figures \ref{fig6}-\ref{fig7}(a)-(b) show the source image and the degraded image for $\sigma_2 = 2$, respectively. The inverted results and the residuals (absolute value) are depicted in figures \ref{fig6}-\ref{fig7}(c)-(d) and figures \ref{fig6}(e)-(f), respectively.
We compare the result of a model that was trained with images of both degradation types (mixed training) with a model that was trained solely for the Gaussian blur with a matching $\sigma_2=2$ (targeted training). As can be visually observed, in accordance with the Double Meaning Theorem, mixed training yields inferior results compared with targeted training where the training matches with the blurring kernel. As can be seen, as expected, the model cannot adhere to both required resolution levels and it ``collapses to the mean''.

\subsection{Example 2 - Sampling Shift}
\begin{figure}[t!]
    \begin{subfigure}[t]{0.165\textwidth}
        \includegraphics[width=0.99\linewidth]{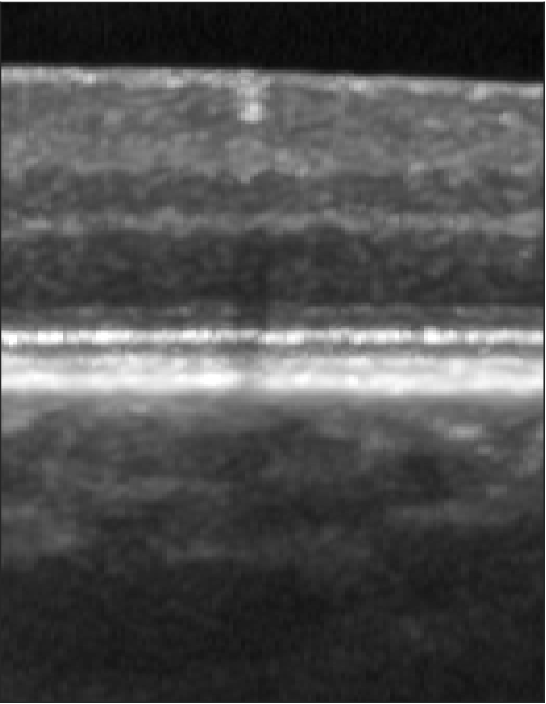}
				\caption{}
    \end{subfigure}%
		\hfill 
		\begin{subfigure}[t]{0.165\textwidth}
        \includegraphics[width=0.99\linewidth]{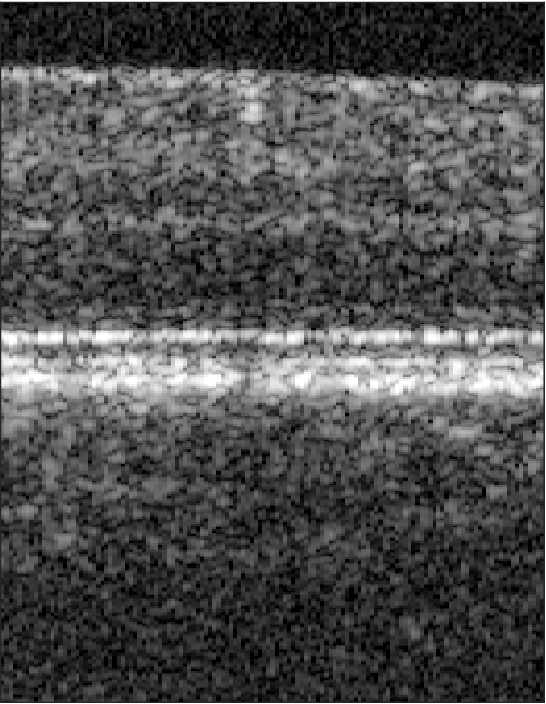}
				\caption{}
    \end{subfigure}%
		\hfill 
    \begin{subfigure}[t]{0.165\textwidth}
        \includegraphics[width=0.99\linewidth]{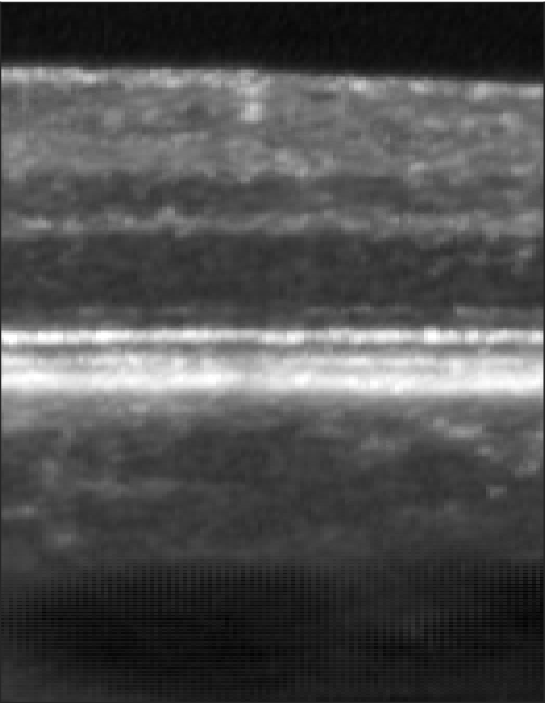}
				\caption{}
    \end{subfigure}%
		\hfill 
		\begin{subfigure}[t]{0.165\textwidth}
        \includegraphics[width=0.99\linewidth]{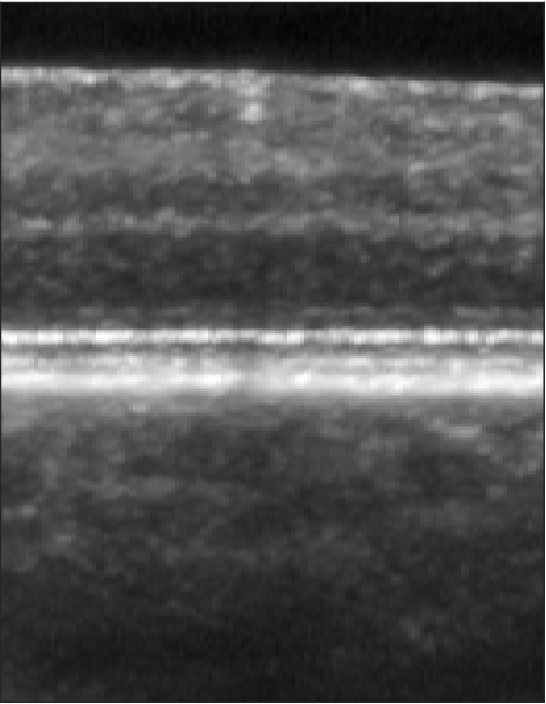}
				\caption{}
    \end{subfigure}%
		\hfill
		\begin{subfigure}[t]{0.165\textwidth}
				\includegraphics[width=0.99\linewidth]{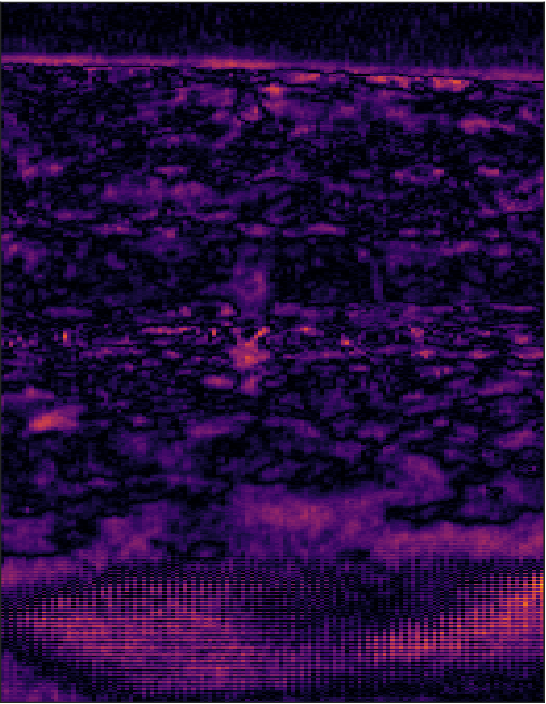}
				\caption{}
		\end{subfigure}%
		\hfill
		\begin{subfigure}[t]{0.165\textwidth}
				\includegraphics[width=0.99\linewidth]{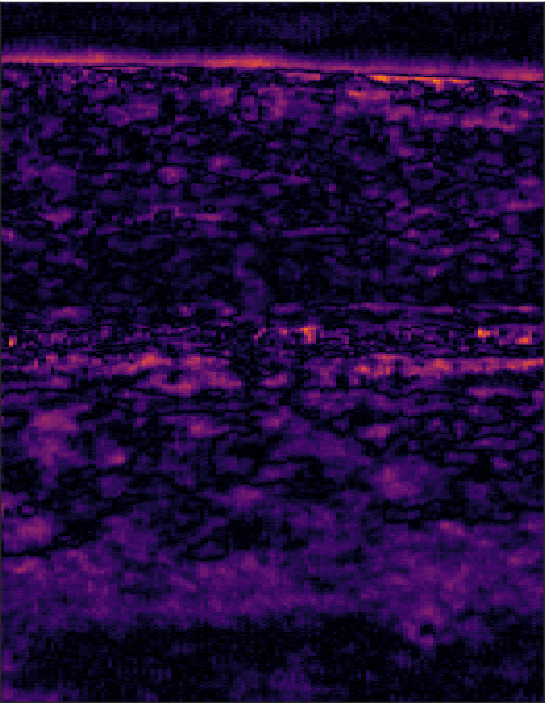}
				\caption{}
    \end{subfigure}%
\caption{{\small Visual comparison of despeckling of OCT retinal image section: (a) Ground truth; (b) Input, 24.64dB; (c) U-Net with mixed training, 33.18dB; (d) U-Net with targeted training, 34.97dB; (e) Residual for mixed training; (f) Residual for targeted training.
No clipping was applied to the images dynamic range throughout processing. Please zoom-in to observe the details.}}
\label{fig8}%
\end{figure}

\begin{figure}[t!]
    \begin{subfigure}[t]{0.165\textwidth}
        \includegraphics[width=0.99\linewidth]{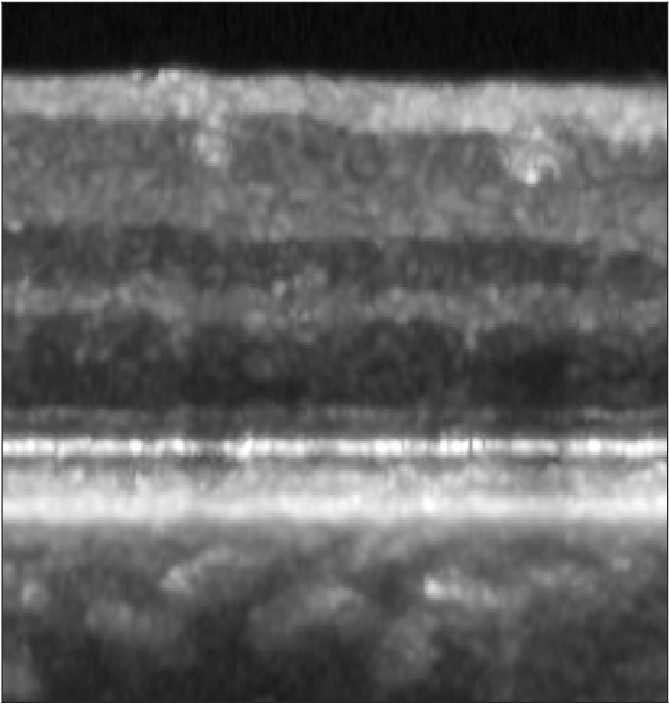}
				\caption{}
    \end{subfigure}%
		\hfill 
		\begin{subfigure}[t]{0.165\textwidth}
        \includegraphics[width=0.99\linewidth]{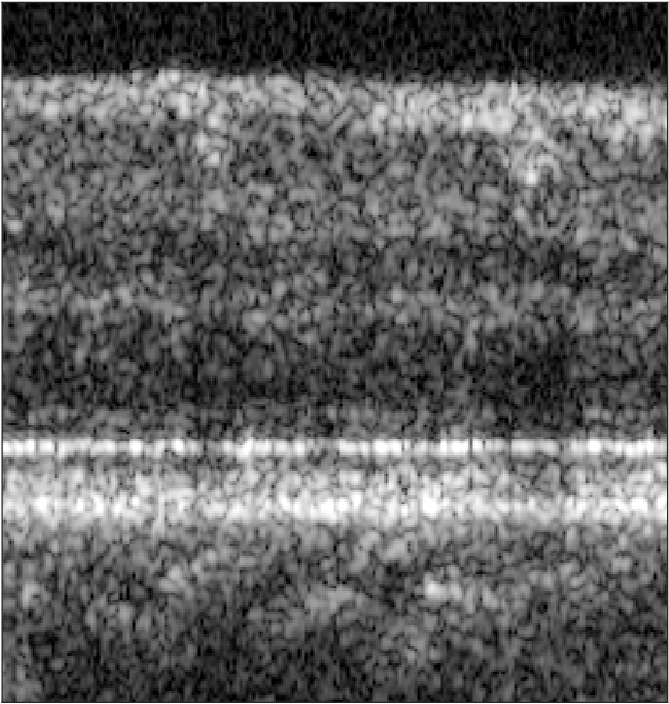}
				\caption{}
    \end{subfigure}%
		\hfill 
    \begin{subfigure}[t]{0.165\textwidth}
        \includegraphics[width=0.99\linewidth]{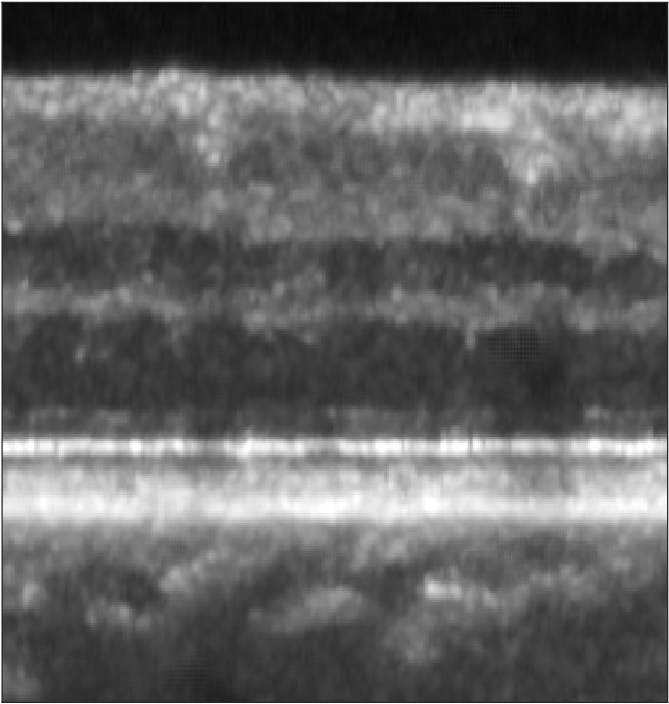}
				\caption{}
    \end{subfigure}%
		\hfill 
		\begin{subfigure}[t]{0.165\textwidth}
        \includegraphics[width=0.99\linewidth]{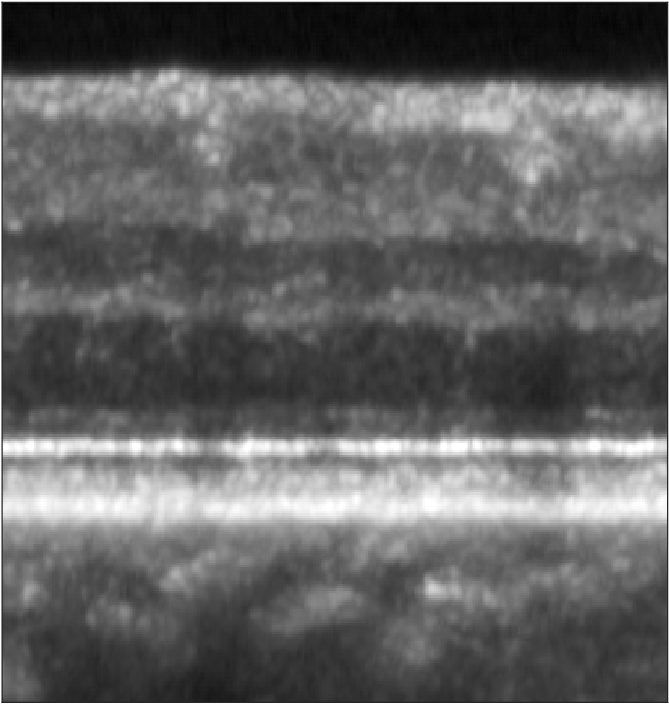}
				\caption{}
    \end{subfigure}%
		\hfill
		\begin{subfigure}[t]{0.165\textwidth}
				\includegraphics[width=0.99\linewidth]{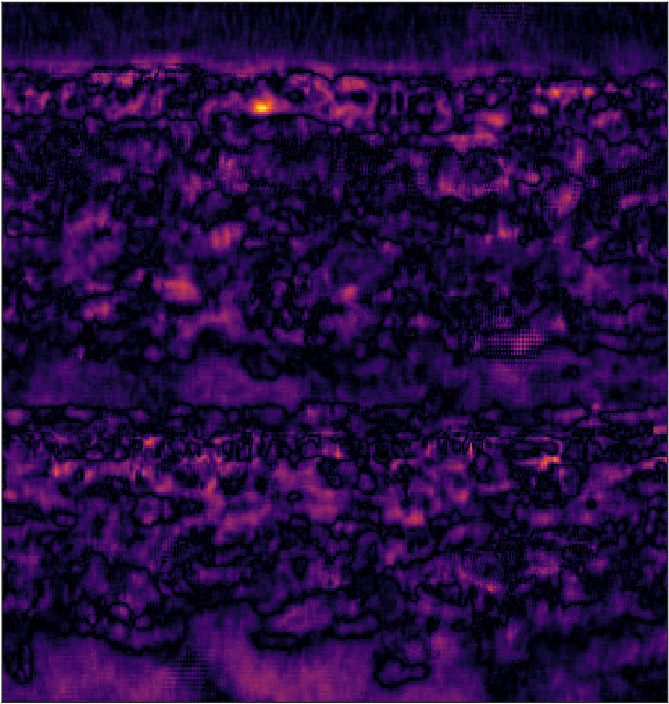}
				\caption{}
		\end{subfigure}%
		\hfill
		\begin{subfigure}[t]{0.165\textwidth}
				\includegraphics[width=0.99\linewidth]{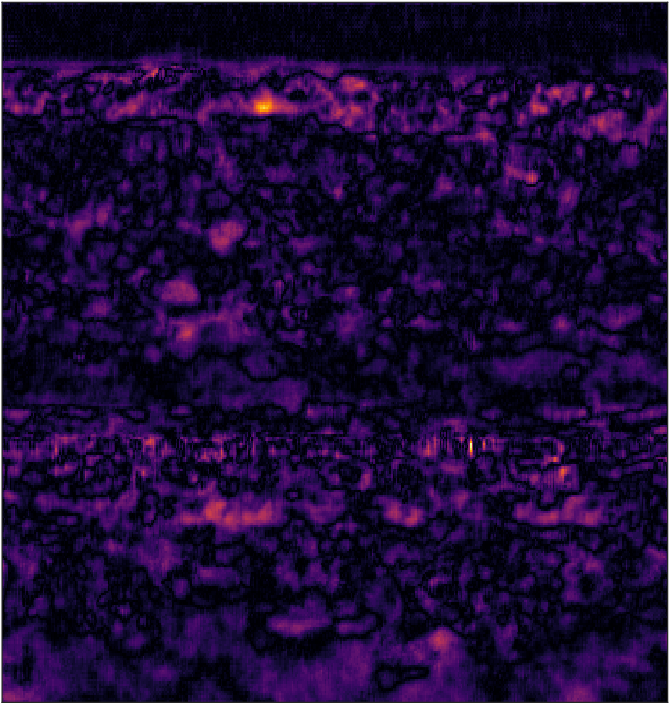}
				\caption{}
    \end{subfigure}%
\caption{{\small Visual comparison of despeckling of OCT retinal image section: (a) Ground truth; (b) Input, 25.35dB; (c) U-Net with mixed training, 32.30 dB; (d) U-Net with targeted training, 35.28dB; (e) Residual for mixed training; (f) Residual for targeted training.}}
\label{fig9}%
\end{figure}

Similarly, we investigate the generalization for speckle-suppression in optical coherence tomography \cite{Pereg:2023B},
and compare the ability of the model to generalize over images of two lateral sampling rates $f^2_{\mathrm{s}}=2f^1_{\mathrm{s}}$. 
We test a model that is trained with retinal images of both sampling rates, and a single matching sampling rate. 
Figures \ref{fig8}-\ref{fig9} show the speckled images, despeckled images obtained by applying Non-Local-Means (NLM) filtering \cite{Cuartas:2018} used as ground truth, the despeckled images produced by the trained models, and the residuals. 
Clearly, mixing sampling rates during training yields degraded results.  
Note the increased error around the blood vessel shadow projections (darker vertical area) in Fig.~\ref{fig8}. 
The blood vessel shadow is almost smeared completely in the mixed training example. 
We observe that, as expected, the models' prediction is more loyal to the measurement than to the NLM ground truth. 
Additionally, we can observe crossed-hatched artifacts as a result of inconsistencies in training sampling rates. 
For these examples the differences between the outputs may often seem subtle, because the averaged terms are visually similar and their average may also seem close to the ground truth.

\subsection{Noisy Training Data}\label{Sec3D}

\paragraph{Noisy Targets}
Let $\Psi=\{\mathbf{y}_{k},\tilde{\mathbf{x}}_{k}\}_{k=1}^m$ be a training set with noisy outputs, where $\tilde{\mathbf{x}_k} \in  \mathcal{X}^{n\times 1}: E[\tilde{\mathbf{x}}|\mathbf{x}]=\mathbf{x}$ are sampled from $p(\tilde{\mathbf{x}}|\mathbf{x})$ and paired with $\mathbf{y}_k \in  \mathcal{Y}^{d\times 1}$ by a deterministic or stochastic mapping as ground truth.
Let $\hat{\mathbf{x}}=\mathcal{F}(\mathbf{y})$ be an inversion model that is trained to 
minimize a cost function,
such that the model $\mathcal{F}$ is trained by minimizing the empirical loss 
\begin{equation}
\mathcal{R}_m(\mathcal{F}) = \frac{1}{m} \sum_{k=1}^m \ell(\mathcal{F}(\mathbf{y}_k),\tilde{\mathbf{x}}_k). 
\end{equation}
where $m$ is the sample size. 
For $m$ large enough, 
\begin{equation}
\hat{\mathbf{x}}= \mathcal{F}(\mathbf{y}) =
\arg\min_{\hat{\mathbf{x}}}
E_{\tilde{\mathbf{x}},\mathbf{y}}\big[\ell(\hat{\mathbf{x}},\tilde{\mathbf{x}})\big],
\end{equation}
For $m \rightarrow \infty$ and a convex loss
\begin{flalign}
\nonumber
\mathcal{R}_m(\mathcal{F}) \rightarrow \mathcal{R}(\mathcal{F})  & = 
E_{\tilde{\mathbf{x}},\mathbf{y}}\big[ \ell(\mathcal{F}(\mathbf{y}),\tilde{\mathbf{x}}) \big] 
= E_{\mathbf{y}} E_{\tilde{\mathbf{x}}|\mathbf{y}}\big[ \ell(\mathcal{F}(\mathbf{y}),\tilde{\mathbf{x}}) \big]
\\
\nonumber
 & 
=
E_{\mathbf{y}} E_{\mathbf{x}|\mathbf{y}}\Big[E_{\tilde{\mathbf{x}}|\mathbf{y}}\big[ \ell(\mathcal{F}(\mathbf{y}),\tilde{\mathbf{x}}|\mathbf{x}) \big] \Big]
\geq 
E_{\mathbf{x},\mathbf{y}}\big[ \ell(\mathcal{F}(\mathbf{y}),\mathbf{x}) \big],
\end{flalign}
which follows from the law of total expectation and Jensen inequality. 

\begin{corollary}
In the special case of MSE loss $\ell(\hat{\mathbf{x}},\tilde{\mathbf{x}})=\|\hat{\mathbf{x}}-\tilde{\mathbf{x}}\|_2^2$, the minimizer is the conditional expectation $\mathcal{F}(\mathbf{y})=E[{\tilde{\mathbf{x}}|\mathbf{y}}]$ \cite{Hastie:2009}. And since the noise is unbiased $E[\tilde{\mathbf{x}}|\mathbf{x}]=\mathbf{x}$, the minimizer is the clean posterior mean 
 $\mathcal{F}(\mathbf{y})=\arg\min_{\hat{\mathbf{x}}}
E_{\mathbf{x},\mathbf{y}}\big[\ell(\hat{\mathbf{x}},\mathbf{x})\big]=E[\mathbf{x}|\mathbf{y}]$. In other words, for MSE loss, unbiased target noise does not affect the learned model. 
\end{corollary}
\begin{proof}
Using the law of total expectation,
\begin{equation*}
E[\tilde{\mathbf{x}}|\mathbf{y}]=E[E[\tilde{\mathbf{x}}|\mathbf{x}]|\mathbf{y}]=E[\mathbf{x}|\mathbf{y}].
\end{equation*}
\end{proof}

\section{Discrete parameter estimation}\label{Sec4}

In the following two sections, our objective is to compare two possible strategies for parameter estimation.
Namely, we compare two Markov chains: 
\begin{enumerate}
\item 
\begin{equation}\label{A}
\theta-X-Y-\hat{\theta}_Y
\end{equation} 
\item 
\begin{equation}\label{B}
\theta-X-Y-\hat{X}-\hat{\theta}_{\hat{X}}
\end{equation}
\end{enumerate}
where $\hat{X}=\mathcal{F}(Y)$, where the mapping $\mathcal{F}: \mathcal{Y}^{d\times 1} \rightarrow \mathcal{X}^{n\times 1}$ is an estimator of $X$ (reconstruction) given the observation $Y$ (or $m$ examples), $\hat{\theta}_Y=g(Y)$ and $\hat{\theta}_{\hat{X}}=f(\hat{X})$ are predictions of $\theta$. 
Note that in some cases the estimation $\hat{X}$ may be obtained from an example (or examples) of $Y|\theta$, that is of a measurement(s) belonging to a specific $\theta$, without the knowledge of $\theta$, but with the knowledge that all these examples originate in the same $\theta$. Alternatively, one could recover $\hat{X}$, from $Y$, based on a recovery method that is general to all $Y$'s regardless of the $\theta$ they originate from. 
The question at the core of this research paper is: Which path leads to a better $\theta$ estimation and when?
Or, in other words, what would be the optimal path towards parameter estimation related with inverse problems? 

We note that $Y$ is not necessarily a deterministic function of $X$. For example, in an inverse problem, $Y$ could be a degraded signal originating in $X$, where the relationship between $X$ and $Y$ could be linear or non-linear, with or without additive noise, such that generally $y=u(x)+v(x)$, where $u(\cdot)$ and $v(\cdot)$ depend on $x$, and are deterministic or random functions. For example, $u(\cdot)$ could represent a blurring process, while $v(\cdot)$ represents an additive noise that could depend on the true value of $x$. 
Additionally, we consider two types of estimators for $\hat{X}$: deterministic and stochastic.
In a continuous setting, given a particular input $Y = y$, a restoration algorithm $\mathcal{F}: \mathcal{Y}^{d\times 1} \rightarrow \mathcal{X}^{n\times 1}$ produces an estimate $\hat{X}$ according to some conditional probability density function $p_{\hat{X}|Y} (\cdot|y)$ such that the estimate $\hat{X}$ is statistically independent of $X$ given $Y$ . For deterministic algorithms,
the conditional probability distribution $P_{X|Y} (\cdot|y)$ is a
delta measure for almost every $y$, while for stochastic
algorithms it is a non-degenerate distribution. 
Furthermore, in many practical cases the estimated distribution $p_{\hat{X}|Y} (\cdot|y)$ is designed as a mapping that is independent of $\theta$.

\subsection{Perfect Perception Estimator}
\begin{definition}[Perfect Perception Recovery]
Let us define a perfect perceptual quality estimator as an estimator that produces an output according to $p_{\hat{X}|Y}(\cdot|y)$ such that
\begin{equation*} 
p_{\hat{X}|Y}(x|y)=p_{X|Y}(x|y).
\end{equation*}
\end{definition}
It has been established in previous work \cite{Blau:2018} that a perfect perceptual quality estimator obeys $p_{\hat{X}}(x)=p_{X}(x)$. In \cite{Blau:2018}, Blau et al. (2018) also showed that perceptual quality and distortion are at odds with each other, and that for the square error distortion $\|\hat{x}-x\|^2$, perfect perceptual quality need not sacrifice more than 3dB in PSNR.

\subsection{Theoretical Analysis}

Consider M classes $\theta_i$ with the conditional probability density functions (pdf's) $p(\mathbf{x}|\theta_i)$ 
defined on the observation space $\mathcal{X}^n$, with priors $P(\theta_i)$, respectively, $i=1,...,M$.
The posterior for $\theta_i$ is 
\begin{equation}\label{6.1}
q_i(\mathbf{x}) \triangleq p(\theta_i|\mathbf{x})=\frac{P(\theta_i)p(\mathbf{x}|\theta_i)}{p(\mathbf{x})},
\end{equation}
where
$p(\mathbf{x})$ is the mixture $p(\mathbf{x})=\sum^M_{i=1}P(\theta_i)p(\mathbf{x}|\theta_i)$.
We note that $0 \leq q_i(\mathbf{x}) \leq 1 \forall \mathbf{x}, \ 1 \leq i \leq M$. 

According to Bayes rule each point $\mathbf{x}$ is assigned to the class $\hat{k}$ with the maximum posterior probability,
$\hat{k} = \mathrm{argmax}_k  \ q_k(x), \ k=1,...,M$ , or equivalently, the minimal loss function \cite{Lissack_1976}.
Namely,%
$\mathbf{x}$ is assigned to class $\theta_i$ if  $\mathbf{x} \in \Gamma_i$,  where 
\begin{equation}\label{6.2}
\Gamma_i= \Big\{ \mathbf{x} \in \mathcal{X}^n \ \Big| \ \sum^M_{k=1} c_{ik} P(\theta_k)p(\mathbf{x}|\theta_k) < \sum^M_{k=1}
c_{lk} P(\theta_k)p(\mathbf{x}|\theta_k), \ l=1,...,M, l \neq i  \Big\},
\end{equation}
where $c_{ij}$ is the loss when a point $\mathbf{x}$ that belongs to class $j$ is assigned the class $i$.
The Bayes minimum risk is formulated as
\begin{equation}\label{6.3}
R^*=\sum_{j=1}^{M}\sum_{i=1}^{M} \int_{\Gamma_i} c_{ij} P(\theta_j)p(\mathbf{x}|\theta_j) d\mathbf{x}
\end{equation}
For $c_{ii}  =  0$  and $c_{ij}  =  1, \ i,  j  =  1,...,M$, 
the Bayes minimum risk is equal to the minimum probability of error $P_e$. 

For the sake of simplicity, let us concentrate on $M=2$. In this case, the Maximum A Posteriori (MAP) classifier operates based on the well known likelihood ratio decision rule,
\begin{flalign}\label{6.3B}
\nonumber
P(\theta_1)p(\mathbf{x}|\theta_1)>P(\theta_2)p(\mathbf{x}|\theta_2) \quad => \quad \mathbf{x} \sim  p_{\theta_1}(\mathbf{x}), \\
P(\theta_1)p(\mathbf{x}|\theta_1)<P(\theta_2)p(\mathbf{x}|\theta_2) \quad => \quad \mathbf{x} \sim  p_{\theta_2}(\mathbf{x}),
\end{flalign}
yielding the decision risk (the outcome expected loss),
\begin{equation}\label{6.4}
R_2^*= P(\theta_2)\int_{\Gamma_1} p(\mathbf{x}|\theta_2)d\mathbf{x} + P(\theta_1)\int_{\Gamma_2} p(\mathbf{x}|\theta_1)d\mathbf{x},
\end{equation}
where,
\begin{equation}\label{6.5}
\Gamma_1= \Big\{ \mathbf{x} \in \mathcal{X}^n \ \Big| \  P(\theta_2)p(\mathbf{x}|\theta_2) < 
P(\theta_1)p(\mathbf{x}|\theta_1) \Big\},
\end{equation}
and
\begin{equation}\label{6.6}
\Gamma_2= \Big\{ \mathbf{x} \in \mathcal{X}^n \ \Big| \  P(\theta_1)p(\mathbf{x}|\theta_1) < 
P(\theta_2)p(\mathbf{x}|\theta_2) \Big\}.
\end{equation}
Lissack et al., 1976 \cite{Lissack_1976} defined a separability measure between two classes, 
\begin{equation}\label{6.7}
J^x_\alpha(\theta_1,\theta_2) = E \phi_\alpha(\mathbf{x}),
\end{equation}
where,
\begin{equation}\label{6.8}
\phi_\alpha(\mathbf{x}) = |q_1(\mathbf{x}) - q_2(\mathbf{x})|^\alpha \quad 0 \leq \alpha \leq \infty.
\end{equation} 
For $\alpha=1$, using $\mathrm{min}(q_1,q_2)= \frac{1}{2}(1-|q_1-q_2|)$ and that $q_1+q_2=1$, it can be shown that the probability of error of the Bayes classifier is 
\begin{equation}\label{6.9}
P_e= \frac{1}{2}[1-J_1(\theta_1,\theta_2)].
\end{equation}
\textit{Proof}
\begin{flalign*}
P_e & = E_x \ \mathrm{min}(q_1(\mathbf{x}),q_2(\mathbf{x})) 
=  \int \mathrm{min}(q_1(\mathbf{x}),q_2(\mathbf{x})) p(\mathbf{x}) d\mathbf{x} \\
& = \frac{1}{2} \int \big( 1- |q_1(\mathbf{x})-q_2(\mathbf{x})| \big) p(\mathbf{x}) d\mathbf{x} 
= \frac{1}{2} [1-J_1(\theta_1,\theta_2)]. \qed
\end{flalign*}
The bounds for $\alpha \neq 1$ and for $M>2$ is provided in Lissack et al., 1976 \cite{Lissack_1976}.
In other words, the probability of misrecognition is inherently bounded by the absolute value of the difference between the posterior pdfs.  

We observe that
\begin{equation}\label{6.10}
J^x_1(\theta_1,\theta_2) = \int \big|P(\theta_2) p(\mathbf{x}|\theta_2) - P(\theta_1) p(\mathbf{x}|\theta_1)\big|d\mathbf{x}.
\end{equation}
Let us denote $P(\theta_1)=\beta$, $P(\theta_2)=1-\beta$, $0 \leq \beta \leq 1$. 
Hence,
\begin{equation}\label{6.11}
J^x_1(\theta_1,\theta_2) = \int \big| \beta p(\mathbf{x}|\theta_1) - (1-\beta) p(\mathbf{x}|\theta_2) \big| d\mathbf{x}.
\end{equation}
For $Y=T(x)$, we have,
\begin{equation}\label{6.12}
J^y_1(\theta_1,\theta_2) = \int \big| \beta p(\mathbf{y}|\theta_1) - (1-\beta) p(\mathbf{y}|\theta_2) \big| d\mathbf{y}.
\end{equation}
Since, under the assumption that $p(\mathbf{y}|\mathbf{x})$ is independent of $\theta$,
\begin{equation}\label{6.13}
p(\mathbf{y}|\theta)=p_\theta(\mathbf{y})= \int p_\theta(\mathbf{x},\mathbf{y}) d\mathbf{x}
= \int p_\theta(\mathbf{x})p(\mathbf{y}|\mathbf{x}) d\mathbf{x},
\end{equation}
we have, 
\begin{flalign}\label{result_B}
J^y_1(\theta_1,\theta_2) 
& =   \int \Big| \beta \int  p(\mathbf{y}|\mathbf{x}) p(\mathbf{x}|\theta_1) d\mathbf{x} -  (1-\beta) \int p(\mathbf{y}|\mathbf{x}) p(\mathbf{x}|\theta_2) d\mathbf{x} \Big|  d\mathbf{y} \\
\nonumber
& =   \int \Big| \int p(\mathbf{y}|\mathbf{x}) \big[ \beta  p(\mathbf{x}|\theta_1) - (1-\beta)  p(\mathbf{x}|\theta_2) \big] d\mathbf{x} \Big|  d\mathbf{y} \\
\nonumber
& \leq   \int \int  p(\mathbf{y}|\mathbf{x}) \Big| \beta  p(\mathbf{x}|\theta_1) - (1-\beta)  p(\mathbf{x}|\theta_2) \Big| d\mathbf{x}  d\mathbf{y} \\
\nonumber
& =  \int \int  p(\mathbf{y}|\mathbf{x}) d\mathbf{y} \Big| \beta  p(\mathbf{x}|\theta_1) - (1-\beta)  p(\mathbf{x}|\theta_2) \Big| d\mathbf{x}   \\
\nonumber
& =  \int \big| \beta p(\mathbf{x}|\theta_1) - (1-\beta) p(\mathbf{x}|\theta_2) \big| d\mathbf{x} 
= J^x_1(\theta_1,\theta_2),
\end{flalign}
where the inequality follows from $\int |g(x)|dx \geq |\int g(x) dx|$, and the last equality follows from $\int p(\mathbf{y}|\mathbf{x}) d\mathbf{y}=1$. 
In other words, \textit{a classifier operating in the $Y$ domain has larger or equal minimal probability of error as the classifier operating in the $X$ domain}. Formally, denote the error of the Bayes classifier operating on the signal $\mathbf{v}$ as $P_e(\mathbf{v})$. Thus, according to our derivation we now have, 
\begin{theorem}[Bayes error bound following signal transformation]\label{theorem5}
For a measurement $Y$ such that $p_\theta(\mathbf{x},\mathbf{y}) = p_\theta(\mathbf{x})p(\mathbf{y}|\mathbf{x})$ the error of the Bayes classifier operating on the measurement obeys
\begin{equation}\label{6.14}
P_e(\mathbf{y}) \geq P_e(\mathbf{x}).
\end{equation}
\end{theorem}

\subsection*{Recovery Estimation Error}

Similarly to (\ref{result_B}), when $p(\hat{\mathbf{x}}|\mathbf{y})$ does not depend on $\theta$, we have,
\begin{equation}\label{result_D}
J^{\hat{\mathbf{x}}}_1(\theta_1,\theta_2) =
\int \Big| \int p(\hat{\mathbf{x}}|\mathbf{y}) \big[ \beta p(\mathbf{y}|\theta_1) d\mathbf{y} -  (1-\beta) p(\mathbf{y}|\theta_2) \big] d\mathbf{y} \Big|  d\hat{\mathbf{x}} 
\leq  J^y_1(\theta_1,\theta_2) 
\leq J^x_1(\theta_1,\theta_2).
\end{equation}
The assumption that $p(\hat{\mathbf{x}}|\mathbf{y})$ does not depend on $\theta$ is not limiting,
because most image restoration algorithms are designed to predict $X$ solely from $Y$, without any prior information about 
$\theta$ as an additional input. 

\begin{theorem}[Bayes error bound following class-agnostic signal recovery]\label{theorem6}
For a measurement $Y$ such that $p_\theta(\mathbf{x},\mathbf{y}) = p_\theta(\mathbf{x})p(\mathbf{y}|\mathbf{x})$, and a reconstruction $\hat{X}$ such that $p(\hat{\mathbf{x}}|\mathbf{y})$ does not depend on $\theta$, the error of the Bayes classifier operating on the reconstruction obeys
\begin{equation}\label{6.16}
P_e(\hat{\mathbf{x}}) \geq P_e(\mathbf{y}) \geq P_e(\mathbf{x}).
\end{equation}
\end{theorem}

As indicated in (\ref{6.10}), provided that the conditional estimated distribution obeys $p_{\hat{X}|\theta}(\mathbf{x}|\theta)=p_{X|\theta}(\mathbf{x}|\theta)$, the minimal probability of error using the recovered signals is equal to the error produced using the measurements $Y$, as summarized in the following theorem. 
\begin{theorem}[Bayes error bound following perfect conditional perception recovery]\label{theorem7}
Given $\hat{X}|\theta$ is a perfect-perception estimator of $X|\theta$, that is, $p_{\hat{X}|Y,\theta}(x|y,\theta)=p_{X|Y,\theta}(x|y,\theta)$,
or equivalently, given an ideally perfect conditional perception recovered signal $\hat{X}$ such that $p_{\hat{X}|\theta}(\mathbf{x}|\theta)=p_{X|\theta}(\mathbf{x}|\theta)$, the error of the Bayes classifier operating on the recovered signal obeys
\begin{equation}\label{6.15}
P_e(\hat{\mathbf{x}}) = P_e(\mathbf{x}).
\end{equation}
\end{theorem}
That said, a perfect recovery of $p_{X|\theta}(\cdot)$ from the measurement $Y$ implicitly requires that the reconstruction system would identify $\theta$ from the measurement, which could normally be obtained in practice only when the measurements are distinguishable, i.e., that there is no overlap between the support of the distributions $p_{Y|\theta}(\mathbf{y}|\theta_1)$ and $p_{Y|\theta}(\mathbf{y}|\theta_2)$ (see also Section~\ref{Sec5}).

If $p_{\hat{X}|\theta}(x|\theta) \neq p_{X|\theta}(x|\theta)$, then we have 
\begin{equation}\label{result_C}
J^{\hat{x}}_1(\theta_1,\theta_2) 
= \int  \Big| \beta \int p(\mathbf{x}|\theta_1) p(\hat{\mathbf{x}}|\mathbf{x},\theta_1)d\mathbf{x}
- (1-\beta) \int p(\mathbf{x}|\theta_2) p(\hat{\mathbf{x}}|\mathbf{x},\theta_2)d\mathbf{x} \Big| d\hat{\mathbf{x}}.
\end{equation}

\textbf{Remark 5.} Consider two classes such that $A_{X}=\big\{\mathbf{x} \in \mathcal{X}^n : \mathbf{x} \sim p_{\theta_1}(\mathbf{x})\big\}$ and $B_{X}=\big\{\mathbf{x} \in \mathcal{X}^n: \mathbf{x} \sim p_{\theta_2}(\mathbf{x})\big\}$ are disjoint sets. Namely, $A_{X} \cap B_{X} = \emptyset$. In this case, $P_e(\mathbf{x})=0$. In other words, by observing $\mathbf{x}$, a Bayes classifier should be able to perfectly detect the class to which $\mathbf{x}$ belongs to. However, we normally assume there is an ambiguity since for the observation $\mathbf{y}$ the sets are no  longer disjoint, i.e., $A_{Y} \cap B_{Y} \neq \emptyset$, where
$A_{Y}=\big\{\mathbf{y} \in \mathcal{Y}^d : \mathbf{y} \sim p_{\theta_1}(\mathbf{y})\big\}$, and $B_{Y}=\big\{\mathbf{y} \in \mathcal{Y}^d : \mathbf{y} \sim p_{\theta_2}(\mathbf{y})\big\}$. 
In our Markov chain $p(\theta, \mathbf{x},\mathbf{y})=p(\mathbf{y})p(\mathbf{x}|\mathbf{y})p(\theta|\mathbf{x})$.
Therefore
\begin{equation}
\arg\max_{\mathbf{x}} p_\theta(\mathbf{x},\mathbf{y}) = \arg\max_{\mathbf{x}} p(\mathbf{x}|\mathbf{y})p(\theta|\mathbf{x})
\end{equation}
Then, in this case, we note that determining $\hat{\mathbf{x}}$ such that $\hat{\mathbf{x}} = \arg\max_{\mathbf{x}} p(\mathbf{y}|\mathbf{x})$ is inherently equivalent to finding $\theta$ directly from $\mathbf{y}$, although may not be accomplished in the same way in practice.

\textbf{Related Work.}
Jalal et al., 2021 \cite{Jalal:2021} state that in light of the ``White Obama" controversy (Menon et al.,
2020b), it has been suggested that reconstruction algorithms are biased because the datasets used for training and inference are unbalanced, thus the obtained models misrepresent the true population distribution. They \cite{Jalal:2021} also claim that reconstruction algorithms that are designed to maximize accuracy are inclined to increase bias. They define proportional representation (PR) such that given a collection of sets of images $\{\mathcal{X}_{\theta_i}\}_{i=1}^K$
where $\theta_i$ is a vector of sensitive attributes and each $\mathcal{X}_{\theta_i}$ represents the group carrying the sensitive attributes $\theta_i$, 
\begin{equation}
P(\hat{\mathbf{x}} \in \mathcal{X}_{\theta_i} ) =  P(\mathbf{x} \in \mathcal{X}_{\theta_i}) \ \forall i.
\end{equation}
In other words, the reconstruction process should not introduce
bias in the distribution for or against any group.
And conditional proportional representation (CPR) such that
\begin{equation}
 P(\hat{\mathbf{x}} \in \mathcal{X}_{\theta_i}|\mathbf{y}) =  P(\mathbf{x} \in \mathcal{X}_{\theta_i}|\mathbf{y}) \ \forall i.
\end{equation}
They show that randomly sampling from the posterior $\hat{\mathbf{x}} \sim p(\mathbf{x}|\mathbf{y})$ achieves CPR and PR. 
Hence, fairness is accomplished because every image that could have led to the measurement could be represented, and not just the
most likely one. Note that by these definitions the reconstruction is stochastic, and not deterministic, and that any reconstruction that falls outside the set $\mathcal{X}_{\theta_i}$ is considered a wrong outcome.
Furthermore, these definitions ignore the distribution $p_{\theta_i}(\mathbf{x})$.
Thus, Ohayon et al. \cite{Ohayon:2024b} propose other measures of fairness to improve the estimation of the distribution of the different groups. However, note that, in our case study, our objective is to estimate $\theta_i$ to the best accuracy. We are not seeking to improve fairness nor to provide a more diverse distribution of the reconstruction $\mathbf{x}$.

\subsection{Examples}

In this subsection we provide illustrative examples to enhance the readers intuitive understanding of the discrete problem setting, and of our proposed theoretical analysis. 

\subsubsection{Naive tree example}

\begin{figure}[t]
\centering
\includegraphics[width=5.2cm, height=4.5cm]{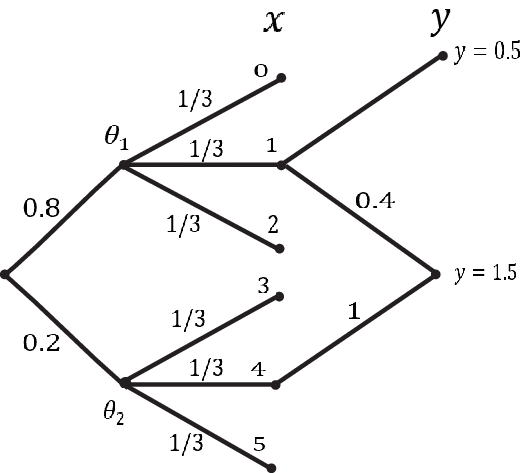}
\caption{Naive tree example}
\label{fig3}
\end{figure}

Consider the setting partially depicted in figure \ref{fig3} such that, $\theta= \{ \theta_1, \theta_2\}$
represent two classes with probabilities $P(\theta_1)=0.8, \ P(\theta_2)=0.2$. The $x$ values in each class are
$A_{X}=\{0,1,2\}$, $B_{X}=\{3,4,5\}$, 
which are conditionally equiprobable, 
$\big\{ x \in A_{X}: p_{\theta_1}(x)=\big[\frac{1}{3},\frac{1}{3},\frac{1}{3}\big] \big\}$,
$\big\{ x \in B_{X}: p_{\theta_2}(x)=\big[\frac{1}{3},\frac{1}{3},\frac{1}{3}\big] \big\}$.
The $y$ values in each class are $A_{Y}=\{0.5,1.5,2.5\}$, $B_{Y}=\{1.5,3.5,4.5\}$,
and the conditional probabilities of observing $y=1.5$ are
$p(y=1.5|x=4)=1, \ p(y=1.5|x=1)=0.4$. 
Namely, the original sets are disjoint but the observation sets partially coincide, 
and one set is 4 times more likely than the other. \\
Assume we observed $y=1.5$, then if we try to derive $\hat{x}$ by maximizing the likelihood 
$\hat{x} = \arg\max_x p(y=1.5|x)$, we get $x \in B_X $, therefore $\theta_{\hat{x}}=\theta_2$.
However, $p(x=4,y=1.5)=0.2 \times 0.33 \times 1 = 0.0667$, $p(x=1,y=1.5)=0.8 \times 0.33 \times 0.4 = 0.1067$.
And since $p(x,y)=p(y)p(x|y)$, we have \\
$p(x \in A_X | y =1.5) > p(x \in B_X | y =1.5)$, then sampling from the posterior is more likely to yield the answer $\theta_{\hat{X}}=\theta_1$. 

\subsubsection{MMSE Estimator and posterior sampling}\label{Sec4.3.2}
We assume a source signal in a discrete setting belonging to two possible classes such that $\big\{ \mathbf{x} \in A_{X}: \mathbf{x} \sim P_{\theta_1}(\mathbf{x}) \big\}$, 
$\big\{ \mathbf{x} \in B_{X}: \mathbf{x} \sim P_{\theta_2}(\mathbf{x}) \big\}$ are \textit{disjoint} sets, i.e., $A_X \cap B_X = \emptyset$, 
where
$P_X(\mathbf{x})=P(\theta_1)P(\mathbf{x}|\theta_1)+P(\theta_2)P(\mathbf{x}|\theta_2)$.
And a degradation process that is a deterministic mapping, i.e., $\big\{\mathbf{y}=T(\mathbf{x}): P(\mathbf{y}|\mathbf{x})=\delta(\mathbf{y}-T(\mathbf{x})), A_y \cap B_y \neq \emptyset \big\}$, where $T: \mathcal{X}^n \rightarrow \mathcal{Y}^d$ is a non-invertible function. In other words, when we observe $\mathbf{y}$ there may be an ambiguity regarding the $\mathbf{x}$ it originates from. Given a noisy observation, the linear least squares estimator (also called "minimum mean squared error" (MMSE) estimator) of the true signal is well known to be the conditional mean of the posterior, 
\begin{equation}
\hat{\mathbf{x}} = E\{X|Y\}= \sum_{\mathbf{x} \ : \ \mathbf{y}=T(\mathbf{x})} \mathbf{x} \ P(\mathbf{x}|\mathbf{y})  = 
\sum_{\mathbf{x} \ : \ \mathbf{y}=T(\mathbf{x})} \mathbf{x} \ \frac{P(\mathbf{y}|\mathbf{x})P(\mathbf{x})}{P(\mathbf{y})}
\end{equation}
For illustration and clarity, let us further assume that there exists an observation $\tilde{\mathbf{y}}$ that is the image of one signal from each class, $\mathbf{x}_A \in A_X$ and of  $\mathbf{x}_B \in B_X$, such that $\tilde{\mathbf{y}}=T(\mathbf{x} \in A_X)$ and $\tilde{\mathbf{y}}=T(\mathbf{x} \in B_X)$, then for that observation we have
\begin{equation}
\hat{\mathbf{x}}(\tilde{\mathbf{y}}) = \mathbf{x}_A\frac{P_X(\mathbf{x}_A)}{P_Y(\tilde{\mathbf{y}}=T(\mathbf{x}_A))} 
+ \mathbf{x}_B \frac{P_X(\mathbf{x}_B)}{P_Y(\tilde{\mathbf{y}}=T(\mathbf{x}_B))}, 
\end{equation}
where $P_X(\mathbf{x}_A)=P(\theta_1)P_{\theta_1}(\mathbf{x}_A)$, and $P_X(\mathbf{x}_B)=P(\theta_2)P_{\theta_2}(\mathbf{x}_B)$,
$P_Y(\mathbf{y})=P(\theta_1)P_{\theta_1}(\mathbf{y})+P(\theta_2)P_{\theta_2}(\mathbf{y})$.

In other words, the reconstruction $\hat{\mathbf{x}}(\tilde{\mathbf{y}})$ is a weighted linear combination of the possible source signals explaining the given measurement. This outcome is also referred to as ``regression to the mean'' that often produces a significantly blurry outcome in image restoration tasks \cite{Elad:2023} (although not accurately representing the original term \cite{Efron:2011}).  In our case study, obviously, the classification of the averaged $\hat{\mathbf{x}}(\tilde{\mathbf{y}})$ is under-determined. If by chance, for example, the averaged image belongs in one of the disjoint sets, then the image will be classified according to that class, regardless of the original image class. Recent work \cite{Ohayon:2024a} suggested to transport the outcome to the distribution of the source signal $P_X(\mathbf{x})$. However, in the case of restoration for the sake of downstream classification, this process essentially determines the estimated class. 
Let us denote,
\begin{equation}
c_A = \frac{P(\theta_1)P_{\theta_1}(\mathbf{x}_A)}{P_Y(\tilde{\mathbf{y}})} , \quad c_B=\frac{P(\theta_2)P_{\theta_2}(\mathbf{x}_B)}{P_Y(\tilde{\mathbf{y}})}.
\end{equation}
We can observe that the ratio is controlled by a similar principle to the likelihood ratio in (\ref{6.3B}). Naturally, the class that has more weight according to the same rule $P(\theta_1)P(\mathbf{x}|\theta_1) \lessgtr P(\theta_2)P(\mathbf{x}|\theta_2)$ is more dominant. Clearly, the fairness of this process is compromised, because when the weight of one class is larger than another, the solution is biased toward that class.%

A similar principle holds when the degradation process is not a deterministic mapping, i.e. 
$P(\mathbf{y}|\mathbf{x})$ is not a delta function, 
\begin{equation}
\hat{\mathbf{x}} = E\{X|Y\}= \sum_{\mathbf{x} \ : \ y=T(\mathbf{x})} \mathbf{x} \ P(\mathbf{x}|\mathbf{y})  = 
\sum_{\mathbf{x} \ : \ \mathbf{y}=T(\mathbf{x})} \mathbf{x} \ \frac{P(\mathbf{y}|\mathbf{x})P(\mathbf{x})}{P(\mathbf{y})}
\end{equation}
Therefore,
\begin{equation}
\hat{\mathbf{x}}(\tilde{\mathbf{y}}) = \mathbf{x}_A\frac{P(\theta_1)P(\mathbf{\tilde{y}}|\mathbf{x}_A)P_{\theta_1}(\mathbf{x}_A)}{P_Y(\tilde{\mathbf{y}}=T(\mathbf{x}_A))} 
+ \mathbf{x}_B \frac{P(\theta_2)P(\mathbf{\tilde{y}}|\mathbf{x}_B)P_{\theta_2}(\mathbf{x}_B))}{P_Y(\tilde{\mathbf{y}}=T(\mathbf{x}_B))}. 
\end{equation}

When $P(\mathbf{y}|\mathbf{x})$ is a stochastic mapping and there are several source signals in each class that can map 
to the same $\mathbf{y}$ we have,
\begin{flalign}
\nonumber
\hat{\mathbf{x}} &= E\{X|Y\}= \sum_{\mathbf{x} \ : \ y=T(\mathbf{x})} \mathbf{x} \ P(\mathbf{x}|\mathbf{y})  = 
\sum_{\mathbf{x} \ : \ \mathbf{y}=T(\mathbf{x})} \mathbf{x} \ \frac{P(\mathbf{y}|\mathbf{x})P(\mathbf{x})}{P(\mathbf{y})}
\\
&=
\sum_{\mathbf{x} \in A_X \ : \ \mathbf{y}=T(\mathbf{x})} \mathbf{x} \ \frac{P(\mathbf{y}|\mathbf{x})P(\mathbf{x})}{P(\mathbf{y})}
+
\sum_{\mathbf{x} \in B_X \ : \ \mathbf{y}=T(\mathbf{x})} \mathbf{x} \ \frac{P(\mathbf{y}|\mathbf{x})P(\mathbf{x})}{P(\mathbf{y})}.
\end{flalign}

In the general case,
\begin{equation}
P(\mathbf{x}|\mathbf{y}) = \sum_i P(\mathbf{x}|\theta_i,\mathbf{y})P(\theta_i|\mathbf{y})= \sum_i P_{\theta_i}(\mathbf{x}|\mathbf{y}) P(\theta_i|\mathbf{y}).
\end{equation}
We observe that the first term in the product $P(\mathbf{x}|\theta_i,\mathbf{y})$ is the posterior for the specific class, in a world where the other classes don't exist. 
The second term in the product,
\begin{equation}
P(\theta_i|\mathbf{y})= q_i(\mathbf{y}) = \frac{P(\theta_i)P(\mathbf{y}|\theta_i)}{P(\mathbf{y})},
\end{equation}
captures the class' probability.
For a continuous parameter,
\begin{equation}
p(\mathbf{x}|\mathbf{y}) = \int p(\mathbf{x}|\theta,\mathbf{y})p(\theta|\mathbf{y}) d\theta= \int p_{\theta}(\mathbf{x}|\mathbf{y}) p(\theta|\mathbf{y}) d\theta.
\end{equation}

\subsubsection{Classification of a degraded/transformed signal}
There are numerous examples of applications where given measurements are first processed to follow with a downstream task of classification, including: audio signals \cite{Rosenbaum:2025}, speech and text \cite{Dissen:2024}, communication \cite{Clancy:2020}, computer vision \cite{Pereg:2024B,Elad:2023},  biomedical imaging \cite{Hanania:2025,Pereg:2023B}, seismic imaging \cite{Pereg:2021,Pereg:2020b} and more. For example, in applications where the signal first goes through an inversion process, decompression, transformation into a different domain (such as: Radon transform) and so on. Then, the restored $\hat{\mathbf{x}}$ is classified to perform a diagnosis, or for other possible uses. 
Alternatively, some propose an end-to-end approach where the measurement is directly processed to perform the classification directly from the measurements. The discrete parameter $\theta$ represents the class to which the signal belongs as described above. 

Note that in (\ref{result_D}) we required that the inversion process $p(\hat{\mathbf{x}}|\mathbf{y})$ does not depend on $\theta$. This assumption is relevant since in many practical examples of real-world classification models that operate on a reconstructed image, the restoration model is agnostic to $\theta$,
as demonstrated in Fig.~\ref{fig1a}. In other words, the restoration model is designed to yield an image that belongs to the general distribution of natural images (biomedical images, or any other specific distribution of signals). This is the case for biomedical signal classification after signal reconstruction/restoration. Namely, the solved inverse problem is essentially implementing a process $p(\hat{\mathbf{x}}|\mathbf{y})$ (rather than $p_{\theta}(\hat{\mathbf{x}}|\mathbf{y})$), because the restoration model at that stage does not ``know"' the class to which the image belongs to. 
Theorem \ref{theorem6} is stating that although the inversion process may yield a visually enhanced image, the information in the measurement about the following classification can only be lost in the prior processing stage.

\subsubsection{Segmentation of noisy (degraded) image}
In many cases, instead of having one label for the entire image (or signal) we are interested in the label of each pixel (or point-value), a segment, or a patch. It is largely hypothesize and demonstrated in experimental results that denoising enhances segmentation results \cite{Ye:2023b,Gruber:2023,Zhao:2023}. Although, as we have shown above, in theory, due to the erroneous nature of the denoising process, equivalent accuracy could be obtained by performing segmentation directly from the noisy measurements.

\section{Cram\'er-Rao Bound for Estimation After Data Processing}\label{Sec5}

In a continuous setting, denote $\Delta_Y=J^{-1}_Y(\theta)$, and $\Delta_{\hat{X}}=J^{-1}_{\hat{X}}(\theta)$, such that by Cram\'er-Rao Bound, for unbiased estimators we have
\begin{equation}
\varepsilon_Y \triangleq E(\theta-\hat{\theta}_Y)^2 \geq \Delta_Y,
\end{equation}
and
\begin{equation}
\varepsilon_{\hat{X}} \triangleq E(\theta-\hat{\theta}_{\hat{X}})^2 \geq \Delta_{\hat{X}},
\end{equation}
By definition of $\hat{X}=\mathcal{F}(Y)$, we have $I(\theta;Y) \geq I(\theta;\hat{X})$, and $J_Y(\theta)\geq J_{\hat{X}}(\theta)$ \cite{Zamir:1998}. Thus, clearly, $\Delta_Y \leq \Delta_{\hat{X}}$. 
That is to say, the estimation error of a parameter $\theta$ of the observed measurement $Y$ is bounded by a smaller or equal quantity comparing with the estimation error of that parameter extracted by any deterministic representation that is obtained via processing of that measurement. 
Equality holds if $\hat{X}$ is a sufficient statistic relative to the family $\big\{ f_\theta(y) \big\}$.
Throughout the analysis we prove assuming $\theta$ is scalar. The generalization to a
vector $\mathbf{\theta}$ is straightforward.

Thus, in the context of our problem here, we ask whether $\hat{\theta}_{\hat{X}}$ is as good as the estimator $\hat{\theta}_{X}$? And the question at hand would be, whether $Y$ is a better estimator of $\theta$, or $\hat{X}$? 
How much do we lose (do we?) from trying to recover $X$ before estimating $\theta$?\\

Note that $Y$ is not termed a ``sufficient statistic" of $X$, because $Y$ is not necessarily a deterministic function of $X$.
For this reason, we define the following properties. 
\begin{definition}[sufficient measurement]%
$Y$ is a sufficient measurement of $X$ if both $X$ and $Y$ have a sufficient statistic of $\theta$.
\end{definition}

To determine whether $Y$ is a sufficient representation\footnote{We distinguish between a sufficient representation and a sufficient statistic, although the term sufficient statistic may have been in occasional use in the literature for random variables that are not necessarily a deterministic function.} relative to $X$ for $\theta$, we focus on the Markov chain $\theta-X-Y$.
We know $I(\theta,X) \geq I(\theta,Y)$ (data processing inequality (DPI)). 
\begin{definition}[sufficient representation]
We say $Y$ is a sufficient representation relative to $X$ for $\theta$ if and only if $I(\theta,X) = I(\theta,Y)$. 
In terms of Fisher information, $J_Y(\theta)=J_{X}(\theta)$, if $J_{X|Y}(\theta)=0$.
e.g., if $Y$ is an invertible degradation of $X$, and/or $X$ is deterministic given Y.
\end{definition}

\begin{theorem}[CRB for perfect perception estimator based on implicit parameter likelihood] \label{theorem2}
If $\hat{X}|\theta$ is a perfect-perception estimator of $X|\theta$, that is, $p_{\hat{X}|Y,\theta}(x|y,\theta)=p_{X|Y,\theta}(x|y,\theta)$,
then, $Y$ is a sufficient representation of $X$, and the CRB for $\hat{\theta}_{\hat{X}}$ and for $\hat{\theta}_{Y}$ is the same. 
\\
In other words, $\varepsilon_{Y} \geq \Delta_Y$, and $\varepsilon_{\hat{X}} \geq \Delta_{\hat{X}}$ where $\Delta_Y=\Delta_{\hat{X}}$.
\end{theorem}
\begin{proof}
Since $p_{\hat{X}|Y,\theta}(x|y,\theta)=p_{X|Y,\theta}(x|y,\theta)$, we have $p_{\hat{X}|\theta}(x|\theta)=p_{X|\theta}(x|\theta)$, 
\begin{equation*}
p_{\hat{X}|\theta}(x|\theta)=\int p_{\hat{X}|Y,\theta}(x|y,\theta)p_{Y|\theta}(y|\theta) dy
= \int p_{X|Y,\theta}(x|y,\theta)p_{Y|\theta}(y|\theta) dy =p_{X|\theta}(x|\theta).
\end{equation*}
Since $p_{\hat{X}|\theta}(x|\theta)=p_{X|\theta}(x|\theta)$, we have $p_{X}(x)=p_{\hat{X}}(x)$, 
\begin{equation*}
p_{\hat{X}}(x)=\int p_{\hat{X}|\theta}(x|\theta)p_\Theta(\theta) d\theta
= \int  p_{X|\theta}(x|\theta)p_\Theta(\theta) d\theta=p_{X}(x).
\end{equation*}
Therefore $I(\theta,X)=I(\theta,\hat{X})$ and $J_X(\theta)=J_{\hat{X}}(\theta)$.
From (\ref{A}) and (\ref{B}) it follows that $I(\theta;X) \geq I(\theta;Y)$ and $J_X(\theta) \geq J_{Y}(\theta)$ (\cite{Zamir:1998}, Section I(6)).
As stated above, due to DPI $H(\theta|\hat{X}=\mathcal{F}(Y)) \geq H(\theta|Y)$, therefore,
$I(\theta,Y) \geq I(\theta,\hat{X})$, which is only possible if $I(\theta;X) = I(\theta;Y) = I(\theta; \hat{X})$. Therefore, $\hat{X}$ is a sufficient statistic for $\theta$, and $J_Y(\theta)=J_{\hat{X}}(\theta)$. Equivalently, $\Delta_Y=\Delta_{\hat{X}}$. 
\end{proof}

\textbf{Remark 6.}
Theorem \ref{theorem2} holds an implicit condition. As stated in the proof, by assumption it follows that $J_X(\theta)=J_Y(\theta)$, which is obeyed
if $J_{X|Y}(\theta)=0$. In our Markov chain, $J_{Y|X}(\theta)=0$, because $p(y|x)$ does not depend on $\theta$. However, a perfect perception estimator $\hat{X}$ enforces a stricter requirement, that $J_X(\theta)=J_Y(\theta)=J_{\hat{X}}(\theta)$. In other words, for $\hat{X}$ to fully capture the information about the parameter $\theta$, $Y$ must be a sufficient representation of $X$. Mathematically speaking, it requires that $J_{X|Y}(\theta)=0$. 
This condition can be satisfied if $p_{X|Y}(x|y)$ does not depend on $\theta$.%
Alternatively, if $p_{X|Y,\theta}(x|y,\theta)$ is deterministic (Given $Y$, $X$ is deterministic), .i.e. $p_{X|Y,\theta}(x|y,\theta)$ is invertible, the FI is perfectly preserved in $Y$, and hence also in $\hat{X}$.

A simple example for an invertible non-deterministic mapping is a noisy constellation $\mathbf{y} = \mathbf{x} + \mathbf{e}$,
where $\mathcal{X}^n = \{\mathbf{x}_1, \mathbf{x}_2, ..., \mathbf{x}_{M} \}$
and $\mathcal{E}^n = \{\mathbf{e}_1, \mathbf{e}_2, ..., \mathbf{e}_{P} \}$ are independent and both uniformly distributed, where we further assume that the map $\mathbf{y} \rightarrow \mathbf{x}$ is surjective. In other words, the noise is stochastic yet the measurements could still be reliably mapped back to their origin. 

\textbf{Remark 7.}
This conclusion may be counter-intuitive since we could have assumed that if we were able to build a generative \textit{stochastic} perfect estimator $p_{\hat{X}|Y}=p_{X|Y}$ then we should be able to estimate $\theta$ with the equivalent or improved accuracy.
Intuitively, as expected by the DPI, it means that even a perfect perception reconstruction $\hat{X}$ will lead to an equivalent bound on the accuracy of the following parameter estimation, \textit{only if the measurement and the source have the same FI}. This condition can be guaranteed when the measurement and/or degradation process did not yield a loss of information that cannot be retrieved with generating samples that are perceptually belonging to the source distribution. 

\textbf{Remark 8.}
Note that the perfect-perception estimator of $X|\theta$ does not necessarily ``know" $\theta$. It may hold it implicitly in a model, e.g., by supervised learning training. 

\begin{theorem}[CRB for perfect perception estimator based on common parameter prior]\label{theorem3}
If $Y$ is a sufficient measurement of $X$, and for a fixed (unknown) $\theta=\theta_k$, $\hat{X}$ is a perfect-perception estimator of $X$, 
that is, $p_{\hat{X}|Y,\theta}(x|y,\theta=\theta_k)=p_{X|Y,\theta,}(x|y,\theta=\theta_k)$,
then, the CRB for $\hat{\theta}_{\hat{X}}$ and for $\hat{\theta}_{Y}$ is the same bound. \\
In other words, $\varepsilon_{Y} \geq \Delta_Y$ and $\varepsilon_{\hat{X}}\geq\Delta_{\hat{X}}$ where $\Delta_Y=\Delta_{\hat{X}}$.
\end{theorem}
\begin{proof}
Since $p_{\hat{X}|Y,\theta}(x|y,\theta=\theta_k)=p_{X|Y,\theta}(x|y,\theta=\theta_k)$ for a fixed (unknown) $\theta=\theta_k$, we have $p_{X|\theta}(x|\theta=\theta_k)=p_{\hat{X}|\theta}(x|\theta=\theta_k)$, since
\begin{equation*}
p_{\hat{X}|\theta}(x|\theta=\theta_k)=\int p_{\hat{X}|Y,\theta}(x|y,\theta=\theta_k) dy
=\int p_{X|Y,\theta}(x|y,\theta=\theta_k) dy =p_{X|\theta}(x|\theta=\theta_k).
\end{equation*}
Therefore $I(X,Y|\theta=\theta_k)=I(\hat{X},Y|\theta=\theta_k)$, and $J_X(\theta)\Big|_{\theta=\theta_k}=J_{\hat{X}}(\theta)\Big|_{\theta=\theta_k}$. 
Consequently, $X-\hat{X}-Y$ forms a Markov chain. Hence $\hat{X}$ is a sufficient statistic for $\theta$, and we have $J_Y(\theta)\Big|_{\theta=\theta_k}=J_{\hat{X}}(\theta)\Big|_{\theta=\theta_k}$.
Equivalently, $\Delta_Y=\Delta_{\hat{X}}$. 
\end{proof}

\begin{proposition}[CRB for perfect prior perception estimator]\label{theorem4}
If $Y$ is a sufficient measurement of $X$, and $\hat{X}$ is a perfect-perception estimator of $X$, that is, $p_{\hat{X}}(x)=p_X(x)$.
Then the CRB for $\hat{\theta}_{\hat{X}}$ and $\hat{\theta}_{Y}$ is not necessarily the same bound. 
In other words, we cannot guarantee that $\Delta_Y=\Delta_{\hat{X}}$.
\end{proposition}
\begin{proof}
$p_{X}(x)=p_{\hat{X}}(x)$ does not necessarily guarantees that $I(\theta,X)=I(\theta,\hat{X})$ nor that $J_X(\theta)=J_{\hat{X}}(\theta)$.
From (\ref{A}) and (\ref{B}) it follows that $I(\theta;X) \geq I(\theta;Y)$.
As stated above, due to DPI $H(\theta|\hat{X}=\mathcal{F}(Y)) \geq H(\theta|Y)$, therefore,
$I(\theta,Y) \geq I(\theta,\hat{X})$.  
\end{proof}

An obvious example is an estimator that produces for every $Y=y$ the same $\hat{X}\in p_X(\cdot)$.
While the estimation clearly belongs to the source distribution, it is obviously quite useless.

\textbf{Remark 9.}
It is important to appreciate the meaning of the above statements and their significance.
It means that, \textit{even when the distribution of $X$ is perfectly recovered, the expected variance of the estimation $\hat{\theta}$ cannot be improved by the perfectly recovered distribution of $X$}. 

\textbf{Remark 10.} Theorems~\ref{theorem2}-\ref{theorem3} imply that the recovery system or method knows or is able to implicitly detect $\theta$, which is in practice, very challenging to accomplish. 

\textbf{Remark 11.}
Recently, generative models have become increasingly popular as restoration strategies in many practical applications. 
Many of these restoration tactics imitate a process leading from $p_Y(y)$ to $p_X(x)$, independent of $\theta$, which is followed by a parameter estimation model. 
The results above emphasize that restoration strategies that are agnostic to $\theta$, i.e., that do not take into consideration the conditional distribution $p_{X|Y,\theta}(x|y,\theta)$, assuming that $X|Y$ is independent of $\theta$, may be erroneous, comparing with a parameter estimation approach that is employed directly on the measurement.
A possible remedy would be to adopt conditional generative processes models. 

An immediate consequence of the DPI is that information about $\theta$ that
is lost in $Y$ cannot be recovered in $\hat{X}$, as $\hat{x}=\mathcal{F}(\hat{y})$ is the outcome of the processing of $Y$.
Namely, it holds that
\begin{equation}
I(\theta;X) \geq I(\theta;Y) \geq I(\theta; \hat{X}), 
\end{equation}
with equality if and only if $\hat{X}$ is a sufficient statistic of $Y$, and $Y$ is a sufficient representation of $X$.
Consequently, the second case we consider is $I(\theta;X) \geq I(\theta;Y) = I(\theta; \hat{X})$.
In this case, $\hat{X}=\mathcal{F}(\hat{Y})$ is a sufficient statistic of $Y$, $X-\hat{X}-Y$ and $\Delta_Y=\Delta_{\hat{X}}$. By Neyman's factorization criterion for a sufficient statistic it holds that $P_{\theta}(Y)=h(Y)Q_{\theta}(\hat{X})$. Note that in this case, regardless of how close $P_{\hat{X}}$ is to $P_X$, the estimation error using the measurement directly or its processing is determined by the distribution of the sufficient statistic relative to $Y$.
Therefore, when $\hat{X}$ is not a sufficient statistic of $Y$, we have our third case, namely $\Delta_Y>\Delta_{\hat{X}}$, since $J_Y(\theta)>J_{\hat{X}}(\theta)$. Table~\ref{Table1} summarizes the above analysis into 3 cases.

\begin{table}[h!]
\centering
\caption{}
\label{Table1}
{\small
\begin{center}
\begin{tabular}{ c | c | c | c }
Estimator & Estimated distribution & Mutual information & CRB \\
\hline
\hline
$p_{\hat{X}|Y,\theta}(x|y,\theta)=p_{X|Y,\theta}(x|y,\theta)$ & $p_{\hat{X}|\theta}(x|\theta)=p_{X|\theta}(x|\theta)$ & $I(\theta;X) = I(\theta;Y) = I(\theta; \hat{X})$ & $\Delta_Y=\Delta_{\hat{X}}$ \\ 
\hline
$p_{\hat{X}|Y,\theta}(x|y,\theta) \neq p_{X|Y,\theta}(x|y,\theta)$ & $ p_{\hat{X}|\theta}(x|\theta) \neq p_{X|\theta}(x|\theta)  $ & $I(\theta;X) \geq I(\theta;Y) = I(\theta; \hat{X})$ & $\Delta_Y\leq\Delta_{\hat{X}}$ \\  
\hline
$p_{\hat{X}|Y,\theta}(x|y,\theta) \neq p_{X|Y,\theta}(x|y,\theta)$ & $ p_{\hat{X}|\theta}(x|\theta) \neq p_{X|\theta}(x|\theta) $ & $I(\theta;X) \geq I(\theta;Y) > I(\theta; \hat{X})$ & $\Delta_Y<\Delta_{\hat{X}}$    
\end{tabular}\label{Table1}
\end{center}
}
\end{table}

\textbf{Remark 12.} Ideally domain adaptation, or domain generalization, could have been accomplished if knowledge of $\theta$ would have been sufficient to ``switch" the mapping $\mathcal{F}_{\theta}: \mathcal{Y}\rightarrow \mathcal{X}$.

\textit{Effect of Perception-Distortion Tradeoff}\\
As mentioned above, it has been shown that the minimal deviation between $p_{X}$ and $p_{\hat{X}}$ that can be obtained by an estimator increases as the mean distortion decreases. In other words, restoration algorithms attempting to minimize a distortion measure inherently contradict an attempt to improve perceptual quality. 
In the context of this work, in accordance with Theorems~\ref{theorem2}-\ref{theorem3}, the accuracy of the parameter estimation is determined by the accuracy in retrieving the conditional source’s distribution with respect to the parameter estimated, rather than successfully minimizing a distortion measure. In other words, \textit{an estimator $\hat{X}$ of $X$ that is designed to minimize a distortion measure rather than find the closest source distribution, is bound to be erroneous in terms of the following  parameter estimation.}
Note that by Theorems~\ref{theorem2}-\ref{theorem3}, an estimator that produces the ``wrong" $\hat{X}$ in terms of the quality measure between $X=x$ and $\hat{X}=\hat{x}$, yet the estimation still belongs to the source conditional distribution $\hat{X}|\theta \in p_{X|\theta}(\cdot)$, would be as reliable as the ``correct" reconstruction. For example, suppose we would like to estimate the degree of elasticity of the retina from an OCT speckled image.  As a first step, we despeckle the image, and our model accidentally produces a despeckled image which does not correspond with the input, yet is an OCT image of a possible retina of a patient corresponding to the same elasticity measure. Therefore the erroneous nature of the reconstruction system w.r.t the quality measure between $X=x$ and $\hat{X}=\hat{x}$ is irrelevant to the following downstream task of estimating $\theta$.    
That said, as pointed out in Theorem~\ref{theorem4}, an estimator that attempts to recover a signal that belongs to the general distribution of the ground-truth $p_X \forall \theta$ (e.g., \cite{Ohayon:2024a}) is not guaranteed to produce a better estimation of the desired parameter from which that ground truth belongs to. Overall, even if the recovered signal belongs to the true conditional ground-truth distribution the accuracy would still be as good as the accuracy obtained by estimating $\theta$ directly from the measurement. 

\section{Examples}

\begin{definition}[Surjective mapping]
A mapping $\mathcal{F}: \mathcal{Y} \rightarrow \mathcal{X}$ is a surjective function, if for every $x$, there is a $y$ such that $x=f(y)$. In other words, every element $x$ is the image of at least one element of $y$. It is not required that $y$ be unique.
\end{definition}

\subsection{Deterministic mappings vs stochatic mappings}

Traditionally, most signal restoration algorithms were deterministic, i.e., the algorithm provides one reconstruction solution for a given degraded input. Mathematically speaking, assume $y=g(x)$ such that $g(\cdot)$ is a deterministic function, and $f(\cdot)$ is a surjective function. 
Therefore, $p(X=f(y)|Y=y)=1$, or equivalently, $p_{X|Y}(x|y)=\delta(x-f(y))$.
Given a particular input $Y = y$, a restoration algorithm produces an estimate $\hat{X}$ according to some distribution 
$p_{\hat{X}|Y} (\cdot|y)$ such that the estimate $\hat{X}$ is statistically
independent of $X$ given $Y$ . For deterministic algorithms,
$p_{\hat{X}|Y} (\cdot|y)$ is a delta function for every $y$, while for stochastic
algorithms it is a non-degenerate distribution.
Recall,
\begin{equation}\label{5.1}
H(\theta|X) = \sum_x p(x)H(\theta|X=x) = -\sum_x p(x) \sum_\theta p(\theta|x) \log p(\theta|x)
= -\sum_{x,\theta} p(x,\theta) \log p(\theta|x).
\end{equation}
Assuming $\theta-Y-X$ forms a Markov chain, 
$P(\theta,X)=P(\theta,X|Y=T(X))$, and $H(\theta|X) \leq H(\theta|Y)$, with equality if $g(\cdot)$ is a bijective function.
Thus, in case that $\mathcal{F}: \mathcal{Y} \rightarrow \mathcal{X}$ is a surjective mapping fully captured by the estimator of $\hat{X}$, i.e., $p_{X|Y}(\cdot|y)=p_{\hat{X}|Y}(\cdot|y)$, as stated in theorems \ref{theorem2}-\ref{theorem3} an algorithm that is able to extract the source signal conditional distribution would yield a parameter estimation result that is as accurate as the actual source and as the measurement. 
The simplest example would be $Y=aX$, where $a>0$ is a constant. 

In the case that $y=g(x)$ where $g: \mathcal{X}^n \rightarrow \mathcal{Y}^d$ is a non-invertible deterministic degradation such that $p_{Y|X}(y|x)=\delta(y-g(x))$, Ohayon et al., 2023 \cite{Ohayon:2023a} proved that when a restoration algorithm enforces both perfect perceptual quality , i.e., $p_{\hat{X}}=p_{X}$
and perfect data fidelity, that is $y=g(\hat{x})$ (equivalent to $p_{Y|\hat{X}}=p_{Y|X}$),
it must be a sampler from the posterior distribution $p_{X|Y}$, and is thus required
to be stochastic.
In other words, these are ill-posed inverse problems in which $X$ cannot be retrieved from $Y$
with zero error ($p_{X|Y} (\cdot|y)$ is not a delta function for
almost every $y$).
Problems such as image colorization, inpainting, single-image super-resolution, JPEG-deblocking, and more all follow this assumed structure.
Thus, in these cases, as indicated in Theorems~\ref{theorem2}-\ref{theorem4}, in order to maintain the same level of accuracy for parameter estimation after the reconstruction, there is need for prior knowledge in the restoration process, such as an oracle of $\theta$, and/or a restoration process that does not depend on $\theta$.  

\subsection{When $\hat{\theta}_Y$ and $\hat{\theta}_{\hat{X}}$ coincide}

\subsubsection*{Additive White Noise}
Let us assume, $Y=X+Z$, where $Z\sim\mathcal{N}(0,\sigma_{\mathrm{n}}^2)$ and $X$ are independent.
Hence $Z$ and $\theta$ are independent. 
That is, $\theta-X-Y$ forms a Markov chain. 
As known, generally speaking, in this case, averaging examples of $Y$ yields $X$, that is, $E[Y]=X$. 

\textbf{Score for AWGN noise.} Suppose we observe a noisy observation of an image, $y = x + z$, where $x\in\mathbb{R}^n$ is the original image drawn from
$p_\theta(x)$, and $z\sim \mathcal{N}(0, \sigma_{\mathrm{n}}^2 I_n)$ is a sample of Gaussian white noise. The observation density $p_\theta(y)$ is
related to the prior $p_\theta(x)$ via marginalization:
\begin{equation}\label{5.1}
p_\theta(y)=p(y|x)p_{\theta}(x)= \int{g(y-x)p_{\theta}(x)}dx. 
\end{equation}
where $g(z)$ is the Gaussian noise distribution. Equation (\ref{5.1}) is in the form of a convolution, and thus
$p_\theta(y)$ is a Gaussian-blurred version of the signal prior, $p_{\theta}(x)$.
In this case, $p_{\theta}(y)$ can be factorized to a multiplication only if $p_{\theta}(x)=\delta(x-x_0)$, where $\delta$ is a Kronecker delta. Alternatively, clearly $p(x,\theta|y)\neq p(x|y)p(\theta|y)$, given $y$, $x$ and $\theta$ are not independent because $p(x|\theta,y) \neq p(x|y)$, therefore clearly $\theta-Y-X$ is not a Markov chain.

Notably, in practice, in a discrete setting, for a specific $\theta$, when the noise level is low enough, 
observing $y$ could be sufficient to determine $x$, that is $p(x=x^*|\theta, y=x^*+z) \approx 1$. 
However, when $\theta$ is unknown, the different $x$'s are not distinguishable, that is $p(x|\theta,y) \neq p(x|y)$. 

Assume $x\sim\mathcal{N}(\mu,\sigma^2_\mathrm{x})$, $\theta=\mu$. We define,
\begin{equation*}
\mathcal{L}(x|\mu)=\mathrm{log} \ p(x|\mu)=-\frac{1}{2}\mathrm{log}(2 \pi \sigma^2_\mathrm{x})-\frac{(x-\mu)^2}{2\sigma_\mathrm{x}^2}.
\end{equation*}
Then we have,
\begin{equation*}
\frac{\partial\mathcal{L}(x|\mu)}{\partial \theta}=\frac{(x-\mu)}{\sigma_\mathrm{x}^2}.
\end{equation*}
Consequently,
\begin{equation*}
J(\theta)=-E\frac{\partial^2\mathcal{L}(x|\mu)}{\partial \mu^2}=\frac{1}{\sigma_\mathrm{x}^2}.
\end{equation*}
Given $m$ observations we have,
\begin{equation*}
J^{-1}_m(\mu)=\frac{\sigma_\mathrm{x}^2}{m}.
\end{equation*}

Now, assume $Y=X+Z$, then $Y\sim\mathcal{N}(\mu,\sigma^2_\mathrm{y})$ such that $\sigma^2_\mathrm{y}=\sigma^2_\mathrm{x}+\sigma^2_\mathrm{z}$. 
Assuming that $\theta=\mu\sim\mathcal{N}(\mu_\mathrm{m},\sigma^2_\mathrm{m})$, we have 
$I(\theta,X)=\frac{1}{2} \mathrm{log} (1+\frac{\sigma^2_\mathrm{m}}{\sigma^2_\mathrm{x}-\sigma^2_\mathrm{m}})$,
and
$I(\theta,Y)$=$\frac{1}{2} \mathrm{log} (1+\frac{\sigma^2_\mathrm{m}}{\sigma^2_\mathrm{y}-\sigma^2_\mathrm{m}})$.
As well as
$J^{-1}_{m,X}(\theta)=\frac{\sigma_\mathrm{x}^2}{m}$ and $J^{-1}_{m,Y}(\theta)=\frac{\sigma_\mathrm{y}^2}{m}$.
At first glance, this may seem as a contradiction, since, obviously $I(\theta,X) > I(\theta,Y)$.
Yet, provided that we would be able to recover $\hat{x}$, such that $p_{\hat{X}|\theta}(x|\theta)=p_{X|\theta}(x|\theta)$, which may be approximately possible in practice for some cases, under sufficiently low noise levels, we should be able to achieve the minimal CRB. Recall, that for a memory-less source we have
\begin{equation}\label{5.2}
I(\mathbf{x},\mathbf{y}) \geq \sum^m_{i=1} I(\mathbf{x}_i,\mathbf{y}_i),
\end{equation}
with equality for $p(\mathbf{x}|\mathbf{y})=\Pi^m_{i=1} p(\mathbf{x}_i|\mathbf{y}_i)$. 
Therefore, in the limit where $m\rightarrow \infty$ we have, $\Delta_Y=\Delta_{\hat{X}} \rightarrow 0$.

Which leads us directly to our next point.
Provided that we have $m_{\mathrm{y}}$ examples of $\mathbf{y}$ and $m_{\mathrm{x}}$ examples of $\hat{\mathbf{x}}$, and that $m_{\mathrm{y}}$ and $m_{\mathrm{x}}$ are finite.
We have $J^{-1}_{m,X}(\theta)=\frac{\sigma_\mathrm{x}^2}{m_{\mathrm{x}}}$ and $J^{-1}_{m,Y}(\theta)=\frac{\sigma_y^2}{m_{\mathrm{y}}}$.
Assuming that we have a perfect perceptual estimator of $X$. That is, that we have been able to reduce the noise in the observation from $\sigma_\mathrm{y}^2$ to $\sigma_\mathrm{x}^2$, then we should be able to recover $\theta$ from $\hat{x}$, unless we created a distortion in the process.

For simplicity, let us assume $\sigma^2_\mathrm{z}=(m-1)\sigma^2_\mathrm{x}$ and a fixed $\theta$ for all examples. In this case, the sufficient statistic for $\theta$ is the sample mean $\hat{\theta}_Y=\bar{Y}=\frac{1}{m} \sum_m Y_m$. On the other hand, for a given $\theta$, $\hat{X}=\frac{1}{m} \sum_m Y_m$ such that $\sigma^2_{\hat{x}}=\frac{\sigma_\mathrm{y}^2}{m}$ is a possible estimation of $X$ reducing the SNR by $\sqrt{m}$.
Therefore, in this case $\hat{\theta}_Y$ and $\hat{\theta}_{\hat{X}}$ coincide and therefore $\varepsilon_{Y}=\Delta_Y=\Delta_{\hat{X}}=\varepsilon_{\hat{X}}$.

An optimal linear estimator for $X$ in the above example would be 
$\hat{x}=ay+b$ such that $a=\frac{\sigma_x}{\sigma_y}$, $b=E[Y](1-\frac{\sigma_\mathrm{x}}{\sigma_\mathrm{y}})$.
In that case, in accordance with Theorem~\ref{theorem3}, clearly $P_{\hat{X}}(x)=P_X(x)$, yet obviously $\forall \theta \ : \ \hat{x} \neq x$. 

\paragraph{Variational autoencoders and variational denoising diffusion models.}
A Gaussian encoder (variance preserving) at time step $t$,
\begin{equation}\label{5.3}
q(\mathbf{x}_t|\mathbf{x}_{t-1})=\mathcal{N}(\sqrt{\alpha_t}\mathbf{x}_{t-1},(1-\sqrt{\alpha_t})\mathbf{I}).
\end{equation}
We are interested in learning conditionals $p_{\phi}(\mathbf{x}_{t-1}|\mathbf{x}_t)$, so that we can simulate new data.
Intuitively, the forward process gradually adds noise to the observation $\mathbf{x}_0$, whereas the generative process
gradually denoises a noisy observation. If we assume $\theta=\mathbf{x}_0$ is the clean image, and $\mathbf{x}=\mathbf{x}_t$ such that $0<t<T$ and $\mathbf{y}=\mathbf{x}_T$ is the noise image. The
generative process approximates an intractable reverse process $p_{\phi}(\mathbf{x}_{t-1}|\mathbf{x}_t)$.
The training loss \cite{Ho:2020} uses KL divergence to directly compare $p_{\phi}(\mathbf{x}_{t-1}|\mathbf{x}_t)$ against forward process posteriors, which are tractable when conditioned on $\mathbf{x}_0$.

\subsection{Sparse Inversion}

Suppose we observe a noisy observation of a signal, $\mathbf{y} = \mathbf{Hx} + \mathbf{z}$, where $\mathbf{x}\in\mathbb{R}^N$ is a sparse vector drawn from
$P_\lambda(\mathbf{x})$, where $P_\lambda(\mathbf{x})=\lambda e^{-\lambda \|\mathbf{x}\|_1}$ such that the parameter
$\lambda$ determines the degree of sparsity of $\mathbf{x}$,%
and $\mathbf{z} \sim \mathcal{N}(0, \sigma_\mathrm{z}^2 I_N)$ is a sample of Gaussian white noise of level. 
The Maximum A Posteriori (MAP) estimator is given by,
\begin{flalign}\label{5.4}
\nonumber
\hat{\mathbf{x}} &= \arg\max_{\mathbf{x} \in \mathbb{R}^N } \quad p_{X|Y}(\mathbf{x}|\mathbf{y}) \\
\nonumber
&= 
\arg\max_{\mathbf{x} \in \mathbb{R}^N } \quad \frac{p_{Y|X}(\mathbf{y}|\mathbf{x}) \ p_{X}(\mathbf{x})}{p_{Y}(\mathbf{y})} \\
&= 
\arg\min_{\mathbf{x} \in \mathbb{R}^N } \quad \log {p_{Y|X}(\mathbf{y}|\mathbf{x})}- \log{p_{X}(\mathbf{x})}.
\end{flalign}
Since in this case $\mathbf{y}\sim\mathcal{N}(\mathbf{Hx},\sigma_{\mathrm{z}}^2 I_N)$, and $P_\lambda(\mathbf{x})=\lambda e^{-\lambda \|\mathbf{x}\|_1}$, we have,
\begin{equation}\label{5.5}
\hat{\mathbf{x}} = \arg\min_{\mathbf{x} \in \mathbb{R}^N } \quad \frac{1}{2\sigma_{\mathrm{z}}^2}
\|\mathbf{y-Hx}\|_2^2-\lambda\|\mathbf{x}\|_1,
\end{equation}
which is considered in the literature as the well known sparse coding $\ell_1$ optimization problem (see \cite{Elad:2010}, Chapter 6). 
The first term is (\ref{5.5}) is the data fidelity term, and the second term is the regularization term.  
The sparse coding recovery estimation is an extensively studied problem \cite{Granda:2013, Granda:2013B,Pereg:2017,Pereg:2021}. In \cite{Pereg:2019} we proved that the bound on the estimation of $\mathbf{x}$ under sufficient separation between adjacent non-zero values (spikes) depends on the noise level. In the absence of noise, the MAP estimation should be perfect. 
However, note that solving (\ref{5.5}) accurately \textit{requires the prior knowledge of $\lambda$}.

For the distribution $P_\lambda(\mathbf{x})=\lambda e^{-\lambda \|\mathbf{x}\|_1}$, the score is
\begin{equation}\label{5.6}
s^m(\lambda) = \frac{\partial}{\partial\lambda}\log p^m_{\lambda}(\mathbf{x})=
\frac{\partial}{\partial\lambda} \big( m\log(\lambda)-\lambda\sum_{i=1}^m\|\mathbf{x}_i\|_1 \big).
\end{equation}
Therefore,
\begin{equation*}
\hat{\lambda}_{X}= \arg\max_{\lambda} \quad s^m(\lambda) = \frac{m}{\sum_{i=1}^m \|\mathbf{x}_i\|_1} .
\end{equation*}

\begin{equation}\label{5.7}
J^m_{X}(\lambda)= m J_{X}(\lambda)= - m E\Bigg\{\Bigg( \frac{\partial^2}{\partial\lambda^2}\log p_{\lambda}(\mathbf{x}) \Bigg)^2 \Bigg|\lambda \Bigg\}=\frac{m}{\lambda^2},
\end{equation}

In other words, hypothetically, if we were to first perfectly denoise the signal such that $\|\hat{\mathbf{x}}\|_1=\|\mathbf{x}\|_1$, assuming that our denoising mechanism is errorless, we have  $\lambda^{-1}_{\hat{X}}=\|\hat{\mathbf{x}}\|_1$ ($m=1$). That said, by Theorem \ref{theorem2} we know that even in that case $\Delta_Y = \Delta_{\hat{X}}$. In other words, if we were to estimate $\lambda$ directly from the measurement, then our expected bound on the parameter estimation error is the same. If we solve the MAP for noisy data the upper-bound for $\hat{\mathbf{x}}$ is given by eq. (\ref{5.10}),(\ref{5.12}) in Appendix \ref{appB}, and it cannot be guaranteed that $\lambda_{\hat{X}}=\lambda$, namely, $\Delta_Y \leq \Delta_{\hat{X}}$.
Interestingly, in the noisy case, denoising the measurement will not improve the following expected estimation error. 

This example emphasizes a common misconception in many signal restoration pipelines. We first restore the signal assuming we know the underlying conditional distribution prior for the original signal, but our restoration model often already relies on the parameter that we wish to estimate in a downstream task. 

Let us assume $x(t)=\delta(t-\theta)$, in other words $p_{X|\theta}$ is a delta function.
Assuming $y=h*x+w$, if we were able to recover $p_{\hat{X}|\theta}(x|\theta)=p_{X|\theta}(x|\theta)$, then we know $\theta_{\hat{X}}=\theta$. In this case we have  $I(\theta;X) = I(\theta;Y) = I(\theta; \hat{X})$. In other words, for the sake of recovering $\theta$ we gain nothing from denoising $y$. It could of course be valuable to us to denoise $y$ for perceptual visual evaluation, and in some cases practically more convenient or possible. Yet, in terms of the accuracy of the parameter recovery - we do not gain from first denoising $y$. A similar formulation can be applied to 2D cases where $\theta$ is the support of any object we wish to detect in an image. The question of bounding the error in this case-study remains an open question that we leave for future work.

\section{Discussion $\&$ Conclusions}
When working on inverse problems in different domains, we tend to assume that the inversion process cleans the data and extracts the relevant information. We often forget that the process could also be inherently biased and inherently inclined to lose information.
This study broadly analyzes parameter estimation and domain shift in inverse problems. 
We introduced general definitions for perfect perceptual estimators in the context of parameter estimation in continuous and discrete settings, and presented theoretical bounds for reliable prediction. 
Observing that domain randomization and data augmentation methods attempt the build models that generalize over several distributions, we highlight a profound vulnerability of this approach in the Double Meaning Theorem (Theorem~\ref{theorem8}), which often leads to mode collapse in signal restoration tasks. Our theoretical findings are confirmed in experimental examples. 
Future work can analyze other specific distributions depending on parameter estimation, such as: Bernoulli distribution, and mixture models e.g., product of experts \cite{Hinton:1999}, experimentally investigate other types of domain changes and adversarial attacks in different applications, and develop coping strategies. 
Our future work will investigate how bias during training affects the classification decision,
and the effect of number of measurements, and sample size.

\appendix

\setcounter{section}{0}

\section{Information-Theory Preliminaries} \label{appA}

\subsection{Ergodic Process}
Generally speaking, when a signal is a function of time (or space), an ergodic signal is a stationary signal whose time estimated moments approximate its statistical ensemble moments. 

\begin{definition}[Mean-Ergodic Processes \cite{Papoulis:1991}]
Considering a continuous signal $x(t)$, where the index $t$ denotes time.
The signal mean is $\eta=E[x(t)]$ and its average is $\eta_T=\frac{1}{2T}\int_{-T}^{T}x(t)dt$.
A mean-ergodic process obeys
\begin{equation}
\lim_{T \rightarrow \infty} \eta_T= \lim_{T \rightarrow \infty} \frac{1}{2T}\int_{-T}^{T}x(t) dt = E[x(t)] = \eta. 
\end{equation}
\end{definition}

\begin{definition}[Autocorrelation-Ergodic Processes \cite{Papoulis:1991}]
A random stationary process $x(t)$ is ergodic in autocorrelation if the time-averaged autocorrelation of a single realization equals the ensemble autocorrelation,
\begin{equation}
\lim_{T \rightarrow \infty} \mathcal{R}_T(\tau)= \lim_{T \rightarrow \infty} \frac{1}{2T}\int_{-T}^{T}x(t)x(t+\tau) dt = E[x(t)x(t+\tau)] = \mathcal{R}(\tau).  
\end{equation}
\end{definition}

\subsection{Efficiency}
Efficiency is defined as the ratio between the probable error of the statistic and that of the most efficient statistic (i.e., that achieves the CRB with equality),
\begin{equation}\label{1.5}
V(\hat{\theta}) = \frac{J^{-1}(\theta)}{\mathrm{Var}(\hat{\theta})},
\end{equation}
where $\mathrm{Var}(\hat{\theta}) = E (\hat{\theta}-E(\hat{\theta}))^2$
denotes the variance of the estimator.
We note that, unbiased estimators obey 
\begin{equation}\label{1.7}
\mathrm{Var}(\hat{\theta})= E (\theta-\hat{\theta})^2.
\end{equation} 
Therefore, $V(\hat{\theta})\leq1$.
An unbiased estimator of a parameter $\theta$ that obtains $V(\hat{\theta})=1$ for all values of the parameter, is called efficient.
Equivalently, the estimator achieves the CRB with equality for all $\theta$.

\subsection{Consistency}
Consider an estimator for $\theta$ with sample size $m$ as a function $\mathcal{F}: \mathcal{X}^m \rightarrow \Theta$.
An estimator $\mathcal{F}(X_1,X_2,...,X_m)$ for $\theta$ is said to be consistent in probability if
$\mathcal{F}(X_1,X_2,...,X_m) \rightarrow \theta$ in probability as $m \rightarrow \infty$.

\subsection{Minimal Sufficient Statistic}

\begin{definition}[sufficient statistic \cite{ThomasCover:2006}] Let $ X \sim p_{\theta}(x)$ and let $T(X)$ be a \textit{deterministic} function. 
We call $T$ a sufficient statistic relative to the family $\{p_\theta(x)\}$ for $\theta$ if $\theta-T-X$ forms a Markov chain
(i.e., given $T$, $\theta$ and $X$ are independent).
\end{definition}
A statistic $T(X)$ is sufficient for underlying parameter $\theta$ if the conditional probability distribution of the data $X$, given the statistic, 
$p(X|T(X))$ does not depend on the parameter $\theta$.
A sufficient statistic extracts all the information available in
$X$ about $\theta$. 
The following theorem states this property:
\begin{theorem}[sufficient statistic \cite{ThomasCover:2006}] Let T be a probabilistic function of X. Then, T is a
sufficient statistic for $\theta$ if and only if $ I(T(X);\theta) = I(X;\theta)$.
\end{theorem}
For a general statistic, we have $J_X(\theta)\geq J_{T(X)}(\theta)$. 
For a sufficient statistic, $J_X(\theta)=J_{T(X)}(\theta)$.

\subsection{Data Processing Inequality}
If $\theta-X-T$ forms a Markov chain, then $I(\theta;X)\geq I(\theta;T(X))$,
with equality if and only if $I(\theta;X|T(X))=0$,
i.e., if and only if $\theta-T-X$ forms a Markov chain. That is to say, that given $T(X)$, $X$ and $\theta$ 
are independent.

We know $I(\theta;X)=H(\theta)-H(\theta|X)$ and $I(\theta;T(X))=H(\theta)-H(\theta|T(X))$, that is, the mutual information $I(\theta;X)$ is the reduction in the uncertainty of $\theta$ due to the knowledge of $X$.
Denote $y=T(x)$, then in a discrete setting,
\begin{equation}
p_Y(y)=\sum_{x:y=T(x)}p_X(x)
\end{equation}
Therefore, $H(Y)\leq H(X)$.
In addition, since conditioning reduces entropy, we have
\begin{equation}
H(\theta|X) = H(\theta|X,T(X)) \leq H(\theta|T(X)),
\end{equation}
with equality if and only if given $T(X)$, $X$ and $\theta$ are independent, in other words, $T(X)$ preserves all the relevant information in $X$, about $\theta$.

\subsection{Fisher-Neyman Factorization Theorem}
For a sufficient statistic
\begin{equation}
p_\theta(x) = h(x)q_{\theta}(T(x)).
\end{equation}
Clearly, in this case,
\begin{equation}
\frac{\partial}{\partial \theta}\log (p_\theta(x)) = \frac{\partial}{\partial \theta}\log(q_{\theta}(T(x))).
\end{equation}
Therefore, $J_X(\theta)=J_{T(X)}(\theta)$.
In other words, $T$ is sufficient if and only if the densities can be written as
products of two factors, one factor that depends on the outcome through $T$
only, and a second factor that is independent of the unknown measure \cite{Halmos:1949}. 
Another way of phrasing this result is to say that $T$ is sufficient if and only if the likelihood
ratio of every pair of measures depends on the outcome through $T$ only.

Of course, one may often speak of a statistic sufficient for some of several parameters. The results in this work can
undoubtedly be extended to treat this concept.

When sufficient statistics exist it has been shown that they will be solutions of the equations of Maximum Likelihood \cite{Fisher:1925}.  

\subsection{The Rao-Blackwell Theorem}
Given an unbiased estimator $f(X)$ and a sufficient statistic $T(X)$, a Rao-Blackwell estimator is formed by conditioning $f(X)$ on the sufficient statistic $T(X)$, i.e., $Ef(X)|T(X)$. This estimator does not depend on $\theta$, and will have a smaller variance or be as efficient as any other unbiased estimator \cite{Blackwell:1947}.
That said, note that not all sufficient statistics lead to efficient estimators that achieve the CRB. The efficiency of an estimator also depends on other factors, such as the specific form of the estimator and the underlying distribution of the data.

\subsection{Estimation uncertainty bounds}

\begin{theorem}[Estimation error and differential entropy \cite{ThomasCover:2006}] 
In a continous setting, for any random variable $X$ and estimator $\hat{X}$,
\begin{equation}
E(X-\hat{X})^2 \geq \frac{1}{2 \pi e} e^{2h(X)},
\end{equation}
with equality if and only if $X$ is Gaussian and $\hat{X}$ is the mean of $X$.
\end{theorem}
$h(X)$ is the differential entropy of a continuous source $x$ with density $p(x)$ 
defined as $h(X) \triangleq -\int p(x) \log p(x) dx $.
For any random variable with variance $\beta$, $h(X)\leq\frac{1}{2}\mathrm{ln}(2\pi e \beta^2)$.

\section{Proof of Theorems \ref{theorem8}}\label{appC}
\begin{proof}
The model $\mathcal{F}$ is trained by empirical risk minimization
\begin{equation}
\mathcal{R}_m(\mathcal{F}) = \frac{1}{m} \sum_{k=1}^m \ell(\mathcal{F}(\mathbf{y}_k),\mathbf{x}_k). 
\end{equation}
where $m$ is the sample size. 
For $m \rightarrow \infty$, $\mathcal{R}_m(\mathcal{F}) \rightarrow \mathcal{R}(\mathcal{F})$,
\begin{equation}
{\small
\mathcal{R}(\mathcal{F})= 
E_{\theta,\mathbf{x},\mathbf{y}}\big[ \ell(\mathcal{F}(\mathbf{y}),\mathbf{x}) \big]
= \frac{1}{M} \sum_{i=1}^{M} E_{\mathbf{x},\mathbf{y}|\theta_i}\big[ \ell(\mathcal{F}(\mathbf{y}),\mathbf{x}) \big]=
E_{\mathbf{y}} \Bigg[ \frac{1}{M} \sum_{i=1}^{M} E_{\mathbf{x}|\mathbf{y},\theta_i}\big[ \ell(\mathcal{F}(\mathbf{y}),\mathbf{x}) \big] \Bigg].}  
\end{equation}
Fix an arbitrary $\mathbf{y}$ in the joint support, since $\mathcal{F}$ operates pointwise on $\mathbf{y}$, minimizing the global risk is equivalent to minimizing the inner term for each $\mathbf{y}$, thus
\begin{equation}
\hat{\mathbf{x}}= \mathcal{F}(\mathbf{y}) =
\arg\min_{\hat{\mathbf{x}}}
\frac{1}{M} \sum_{i=1}^{M}E_{\mathbf{x}_i\sim\mathbf{x}|\mathbf{y},\theta_i}\big[\ell(\hat{\mathbf{x}},\mathbf{x}_i)\big],
\end{equation}
The theorem formalizes that domain randomization with uniform weighting induces a conditional Bayes estimator under the mixture posterior across domains.
If $p(\mathbf{x}|\mathbf{y},\theta_i)$ collapses to a Dirac delta at $\mathbf{x}_i$ for each domain $i$, then 
$E_{\mathbf{x}_i\sim\mathbf{x}|\mathbf{y},\theta_i}\big[\ell(\hat{\mathbf{x}},\mathbf{x}_i)\big]=\ell(\hat{\mathbf{x}},\mathbf{x}_i)$
and we have
\begin{equation}
\hat{\mathbf{x}}= \mathcal{F}(\mathbf{y}) =
\arg\min_{\hat{\mathbf{x}}}
\frac{1}{M} \sum_{i=1}^{M}{\ell(\hat{\mathbf{x}},\mathbf{x}_i)}.
\end{equation}
\end{proof}

\section{Sparse Recovery Method and Recovery-Error Bound} \label{appB}
We assume a sparse signal in a discrete setting 
\begin{equation}\label{5.8}
	x[k]=\sum_m{c_m\delta[k-k_m]}, \qquad k \in \mathbb{Z}, \qquad c_m \in \mathbb{R}
\end{equation}
where $\delta[k]$ denotes the Kronecker delta function (see \cite{Ricker:1940}), and $\sum_m|c_m|<\infty$.
$K=\{k_m\}$ is the set of discrete delays corresponding to the spikes locations.
In a noisy environment we consider a discrete signal of the form 
\begin{equation}\label{5.9}
	y[k]=\sum_n{x[n]g_{\sigma,n}[k-n]}+w[k], \qquad n \in \mathbb{Z}
\end{equation}
where $w[k]$ is additive noise, with bounded energy
$
			\|\mathbf{w}\|_1=\|\mathbf{y}-G\mathbf{x}\|_1 \leq \delta,
$
where $\|\mathbf{x}\|_1:=\sum_k |x[k]|$ and $G$ as an operator matrix such that $(G)_{k,n}=g_{\sigma,n}[k-n]$, representing a set of admissible kernels
with parameters $\beta, \varepsilon$ as defined in \cite[Definition 2.1 ]{Pereg:2017}, and a scaling parameter $\sigma$. We denote $F_\mathrm{s}$ as the sampling rate.
The objective is to estimate the true support $K=\{k_m\}$ and the spikes' amplitudes $\{c_m\}$ from the observed signal (e.g., seismic trace) $y[k]$.
In \cite{Pereg:2017} it is shown that the solution $\hat{\mathbf{x}}$ of 
\begin{align}\label{5.10}
{\setstretch{2.25}
	\begin{array}{lll}
&\min 										& \|\mathbf{\hat{x}}\|_1  \\
& {\text{subject to}}			& \|\mathbf{y}-G\mathbf{\hat{x}}\|_1\leq  \delta,
	\end{array}
}
\end{align}
satisfies
\begin{equation}\label{5.11}
 \|\hat{\mathbf{x}}-\mathbf{x}\|_1 \leq 
\frac{4\rho}{\beta \gamma_0}\delta.
\end{equation}
\begin{equation*}
		 \rho\ \triangleq \max{\Big\{\frac{\gamma_0}{\varepsilon^2}, {(F_\mathrm{s}\sigma)^2\alpha_0} \Big\}}
\end{equation*}
\begin{equation*}
\alpha_0=\max\limits_n{g_{\sigma,n}(0)}, \quad  \gamma_0=\min\limits_n{g_{\sigma,n}(0)}.
\end{equation*}
The dependance of $x$ on the time $k$ is not written for simplicity.\\
By assumption eq. (\ref{5.10}) defines a MAP solution where it is assumed that both the solution $\mathbf{x}$ and the noise $\mathbf{w}$ are i.i.d with Laplace zero-mean distribution. 
Following the same proof outline in \cite{Pereg:2017} it is possible to show \cite{Pereg:2019} that
the solution $\hat{\mathbf{x}}$ of 
\begin{align}\label{5.12}
{\setstretch{2.25}
	\begin{array}{lll}
&\min 										& \|\mathbf{\hat{x}}\|_1  \\
& {\text{subject to}}			& \|\mathbf{y}-G\mathbf{\hat{x}}\|_2\leq S_{\mathrm{w}},
	\end{array}
}
\end{align}
satisfies
\begin{equation}
 \|\hat{\mathbf{x}}-\mathbf{x}\|_2 \leq 
\frac{64N\rho^2}{\beta^2 \gamma^2_0}S_{\mathrm{w}}.
\end{equation}
Similarly, eq. (\ref{5.8}) defines a MAP solution where it is assumed that both the solution $\mathbf{x}$ follows a Laplace zero-mean distribution, and the noise is Gaussian i.i.d noise with bounded energy defined by $S_\mathrm{w}$, where 
$\|\mathbf{w}\|_2 \triangleq \sqrt{\sum_k w^2[k]}$.

\bibliography{dpereg_CRB}

\end{document}